\documentclass[12pt,aps]{revtex4}
\usepackage{epsfig,amsmath,amssymb}

\newcommand{\eq}{{ \equiv }}
\newcommand{\fr}[1]{
             \frac{#1}}
\newcommand{\bea}{\begin{eqnarray}}
\newcommand{\eea}{\end{eqnarray}}

\newcommand{\chibar}{\overline{\chi}}

\newcommand{\ket}{{\rangle}}
\newcommand{\e}{{\varepsilon}}
\newcommand{\bra}{{\langle}}
\newcommand{\gc}{\bra\fr{\alpha_s}{\pi}G^2\ket}

\newcommand{\qc}{\bra\,\overline{q}q\,\ket}
\newcommand{\ga}{g_{{\cal A}}}
\newcommand{\gat}{\tilde{g}_{{\cal A}}}
\newcommand{\ebar}{{\bar{\varepsilon}}}
\def\slash#1{#1 \hskip -0.5em / }

\begin{document}
\title{A Heavy-Light Chiral Quark Model}
\author{A.~Hiorth}
\email{aksel.hiorth@fys.uio.no}
\affiliation{Department of Physics, University of Oslo,
P.O.Box 1048 Blindern, N-0316 Oslo, Norway}
\author{J.~O.~Eeg}
\email{j.o.eeg@fys.uio.no}
\affiliation{Department of Physics, University of Oslo,
P.O.Box 1048 Blindern, N-0316 Oslo, Norway}

\begin{abstract}
We present a new chiral quark model for mesons involving a heavy and
 a light (anti-) quark.
The model relates various combinations of a quark - meson coupling constant 
and loop integrals to physical quantities. Then, some quantities may 
be predicted and some used as input.
 The extension from other similar
 models is that the present model
 includes the lowest order gluon condensate of the order
(300 MeV$)^4$ determined by the mass splitting of the $0^-$ and the $1^-$
heavy meson states. Within the model, we find a reasonable
 description of parameters such as
the decay constants $f_B$ and $f_D$, the Isgur-Wise function and the
 axial vector coupling $g_A$ in chiral perturbation theory for light and heavy mesons.
 
\end{abstract}
\maketitle

\section{Introduction}
While the short distance (SD) effects in hadronic  physics are well
 understood within perturbative quantum chromodynamics (pQCD),
 long distance (LD) effects
 have been hard to pin down. Lattice gauge theory should  be
 able to solve the problem,
but the calculations are often very difficult to 
perform.
QCD sum rules might also give the answer in some cases.
Still, in some cases it has been fruitful to use various
 QCD inspired models and assumptions. In the light quark sector,
low energy quantities have been studied in terms of the (extended) Nambu-Jona-Lasinio model (NJL)\cite{bijnes}
and   the chiral quark model ($\chi$QM)\cite{chiqm}, which is the mean
 field approximation of NJL. 

In the $\chi$QM, 
the light quarks ($u, d, s$) couple to the would be Goldstone octet
mesons ($K, \pi, \eta$) in a chiral invariant way, such that all effects 
are in principle calculable in terms of physical quantities and a few
model dependent parameters, namely the quark condensate, the gluon condensate
and the constituent quark mass \cite{pider,epb,BEF}. More specific, one 
may calculate the coupling constants of chiral Lagrangians in this way.

In this paper we will extend the ideas of the chiral quark model
of the pure light sector \cite{chiqm,pider,epb,BEF} to the sector 
involving a heavy quark ($c$ or $b$) and thereby a heavy meson. Such ideas
 have already been presented in previous papers \cite{barhi,effr,itCQM,ebert2}.
 Also in this case, one may calculate the parameters
 of chiral Lagrangian terms,
where the description of  heavy mesons are in accordance with heavy quark 
effective field theory (HQEFT) .
 In the present paper we will extend the ideas of
\cite{barhi,effr,itCQM,ebert2}  to include gluon (vacuum) condensates.
 The 
motivation for the inclusion of gluon condensates is that this works  well
 in order to understand  the $\Delta I  =   1/2$ rule for 
$K \rightarrow 2 \pi$
within the $\chi$QM \cite{pider,BEF} and within generalized 
factorization \cite{cheng}. Furthermore, it allows us to consider effects
related to the gluonic aspect of $\eta'$ as considered in \cite{EHP},
some aspects of $D$-meson decays \cite{EFZ}, and also to calculate
gluon condensate contributions to $B-\bar{B}$-mixing \cite{AHJOE}.

Having established our heavy - light chiral quark model (HL$\chi$QM),
 we can integrate out
the light and heavy quarks and  obtain chiral Lagrangians involving
light and heavy mesons \cite{wise1,itchpt}. Chiral perturbation theory 
($\chi$PT) based
on such Lagrangians works in the pure strong sector. 
In order to define the model and its parameters, we have to integrate out
the quarks, and we will find some typical divergent loop integrals.
We will relate all
divergent loop integrals to some physical parameters, as was 
done in \cite{BEF}. This means in particular that we will treat
quadratic, linear, and logarithmic divergences as different.
 If we need to
calculate a divergent integral we will do so in dimensional
regularization, although various regularization procedures might be used.
Thus, the  regularizing prescription for divergent diagrams
is to be regarded as a part of the model. Still, even if integrals are 
divergent, the effective UV cut-off scale is, as for $\chi$PT,
 the chiral symmetry breaking scale $\Lambda_\chi \simeq$ 1 GeV, where also the
 matching of pQCD and HL$\chi$QM is performed. This is also considered a part 
of our model construction.

The paper is organized as follows: In the next section (\ref{sec:HCQM})
  we describe the 
HL$\chi$QM, and in section \ref{sec:strong}  
 we consider bosonization in  the strong sector and of the weak
 current respectively. In section \ref{sec:param}  we discuss the relations between
 physical and model dependent parameters. In section \ref{sec:iw} we discuss the 
Isgur-Wise function, and in section \ref{sec:mq} we bosonize the $1/m_Q$ terms.
Section VII contains the presentation of some necessary chiral corrections
for our numerical analysis.
Finally in section \ref{sec:chiral} we discuss our results.
Loop integrals are listed in Appendix \ref{app:loop}.  In 
Appendix \ref{app:tranf} we list some 
transformation properties of the involved fields. 

\section{The Heavy -  Light Chiral Quark Model (HL$\chi$QM)}\label{sec:HCQM}

Our starting point is the following Lagrangian containing both quark
 and meson fields:
\begin{equation}
{\cal L} =  {\cal L}_{HQEFT} +  {\cal L}_{\chi QM}  +   {\cal L}_{Int} \; ,
\label{totlag}
\end{equation}
where \cite{neu}
\begin{equation}
{\cal L}_{HQEFT} =  \overline{Q_{v}} \, i v \cdot D \, Q_{v} 
 + \frac{1}{2 m_Q}\overline{Q_{v}} \, 
\left( - C_M \frac{g_s}{2}\sigma \cdot G
 \, +   \, (i D_\perp)_{\text{eff}}^2  \right) \, Q_{v}
 + {\cal O}(m_Q^{- 2})
\label{LHQEFT}
\end{equation}
is the Lagrangian for heavy quark effective field theory (HQEFT).
The heavy quark field  $Q_v$
annihilates  a heavy quark  with velocity $v$ and mass
$m_Q$. Moreover,  
$D_\mu$ is the covariant derivative containing the gluon field
(eventually also the photon field), and 
$\sigma \cdot G = \sigma^{\mu \nu} G^a_{\mu \nu} t^a$, where 
$\sigma^{\mu \nu}= i [\gamma^\mu, \gamma^\nu]/2$, $G^a_{\mu \nu}$
is the gluonic field tensors, and $t^a$ are the colour matrices. 
This chromo-magnetic term has a factor $C_M$, being one at tree level,
 but slightly modified by perturbative QCD.(When the covariant derivative also contains the photon field, there is also a corresponding magnetic term
$\sim \sigma \cdot F$, where $F^{\mu \nu}$ is the electromagnetic tensor).
 Furthermore,  
$(i D_\perp)_{\text{eff}}^2 =
C_D (i D)^2 - C_K (i v \cdot D)^2 $. At tree level, $C_D = C_K = 1$.
Here, $C_D$ is not modified by perturbative QCD, while $C_K$ is different 
from one due to perturbative QCD corrections \cite{GriFa}.

The light quark sector is described by the chiral quark model ($\chi$QM),
having a standard QCD term and a term describing interactions between
quarks and  (Goldstone) mesons: 
\begin{equation}
{\cal L}_{\chi QM} =   \bar{q}(i\gamma^\mu D_\mu  -  {\cal M}_q) q
  -     m(\bar{q}_R \Sigma^{\dagger} q_L   +    \bar{q}_L \Sigma q_R) \; , 
\label{chqmU}
\end{equation}
where $q^T  =  (u,d,s)$ are the light quark fields. The left- and
 right-handed
 projections $q_L$ and $q_R$ are transforming after $SU(3)_L$ and $SU(3)_R$
respectively. ${\cal M}_q$ is the the current quark mass matrix,
$m$ is the ($SU(3)$ -  invariant) constituent quark mass for light quarks, and 
$\Sigma  =   \exp(i \sum_j \lambda^j \pi^j)$ is a 3 by 3 matrix containing
 the (would be)  Goldstone octet ($\pi, K, \eta$) :
\begin{equation}
\xi=e^{i\Pi/f}\, \quad
 \text{where}\, \quad \Pi=\fr{\lambda^a}{2}\phi^a(x) = 
\frac{1}{\sqrt{2}} \left[\begin{array}{ccc} \fr{\pi^0}{\sqrt{2}}+\fr{\eta_8}{\sqrt{6}} & \pi^+
&K^+\\ \pi^-&-\fr{\pi^0}{\sqrt{2}}+\fr{\eta_8}{\sqrt{6}} & K^0\\
K^- &\overline{K^0}& -\fr{2}{\sqrt{6}}\eta_8\end{array}\right] \; .
\end{equation}
 The $\chi$QM has a ``rotated version'' 
with  flavour rotated quark fields $\chi$ given by:
\begin{equation}
\chi_L  =   \xi^\dagger q_L \quad ; \qquad \chi_R  =   \xi q_R \quad ; \qquad 
\xi \cdot \xi  =   \Sigma \; .
\label{rot}
\end{equation}
In the rotated version, the chiral interactions are rotated  into the
kinetic term while the interaction term proportional to $m$ in 
(\ref{chqmU}) 
 become a pure (constituent) mass term \cite{chiqm,BEF}:
\begin{equation}
{\cal L}_{\chi QM} =  
\chibar \left[\gamma^\mu (i D_\mu   +    {\cal V}_{\mu}  +  
\gamma_5  {\cal A}_{\mu})    -    m \right]\chi 
  -     \chibar \widetilde{M_q} \chi \;  , 
\label{chqmR}
\end{equation}
where the vector and axial vector fields 
${\cal V}_{\mu}$ and  
${\cal A}_\mu$ are given by:
\begin{equation}
{\cal V}_{\mu}\eq \fr{i}{2}(\xi^\dagger\partial_\mu\xi
+\xi\partial_\mu\xi^\dagger 
) \qquad ;  \qquad  
{\cal A}_\mu\eq  -  \fr{i}{2}
(\xi^\dagger\partial_\mu\xi
-\xi\partial_\mu\xi^\dagger) \; ,
\label{defVA}
\end{equation}
and  $\widetilde{M_q}$ defines  the rotated version of the current
 mass term:
\bea
&&\widetilde{M_q} \eq \widetilde{M}_q^V   +    \widetilde{M}_q^A \gamma_5  \;
 , \; \text{where} \label{cmass}\\
\widetilde{M}_q^V \, \eq \, 
\fr{1}{2}(\xi^\dagger {\cal M}_q^\dagger\xi^\dagger \,
+&& \xi {\cal M}_q\xi )\quad\text{and}\quad 
\widetilde{M}_q^A\eq -\fr{1}{2}(\xi^\dagger {\cal M}_q^\dagger\xi^\dagger
 -\xi {\cal M}_q\xi) \; .
\label{masst}
\eea
Here $L$ is the left -  handed projector in Dirac space, 
$L =  (1 -  \gamma_5)/2$
and $R$ is the corresponding right -  handed projector, $R =  (1 +  \gamma_5)/2$ .
The Lagrangian (\ref{chqmR}) is manifest invariant under the  unbroken
symmetry $SU(3)_V$.
In the light -  sector, the various pieces of the strong Lagrangian
can be obtained by integrating out the constituent quark fields $\chi$,
and these pieces can be written in terms of the  fields ${\cal A}_\mu \, ,
\, \widetilde{M}_q^V$ and $\widetilde{M}_q^A$ which are manifest invariant
under local $SU(3)_V$ transformations. For instance, the standard
 ${\cal{O}}(p^2)$ kinetic term may (up to a constant) be written as 
$Tr\left[ {\cal{A}}^\mu \, {\cal{A}}_\mu \right]$.
This is easily seen by using the relations
\begin{equation}
{\cal{A}}_\mu \; =  \;  -   \fr{1}{2 i} \xi \, (D_\mu \Sigma^\dagger) \, \xi
\; = +\; \fr{1}{2 i} \xi^\dagger \, (D_\mu \Sigma) \, \xi^\dagger \; ,
\label{ASigma}
\end{equation} 
where $D_\mu$ is a covariant derivative containing the photon field.
 The vector field
${\cal{V}}^\mu$ transforms as a gauge field under local $SU(3)$,
 and can only appear in  combination with a derivative as
a covariant derivative $( i \partial^\mu  +   {\cal{V}}^\mu)$.
   
In the heavy -  light case, the generalization of the
 meson -  quark interactions in the pure light sector  $\chi$QM
is given by the following $SU(3)$ 
invariant Lagrangian:
\begin{equation}
{\cal L}_{Int}  =   
 -   G_H \, \left[ \chibar_a \, \overline{H_{v}^{a}} \, Q_{v} \,
  +     \overline{Q_{v}} \, H_{v}^{a} \, \chi_a \right]   +   
 \fr{1}{2G_3}Tr\left[ \overline{H_{v}^{a}}\,  H_{v}^{a}\right] \; ,
\label{Int}
\end{equation}
where $G_H$ and $G_3$ are  coupling constants, and
 $H_{v}^{a}$ is the heavy meson field  containing
 a spin zero and spin one boson:
\begin{eqnarray}
&H_{v}^{a} & \eq  P_{+} (P_{\mu}^{a} \gamma^\mu -     
i P_{5}^{a} \gamma_5)\; , \nonumber \\
&\overline{H_{v}^a}
& =  \gamma^0 (H_{v}^a)^\dagger \gamma^0
 =  \left[(P_{\mu}^{a})^{\dagger} \gamma^\mu 
 -   i (P_{5}^{a})^\dagger \gamma_5\right] P_{+} \; , \label{barH}
\end{eqnarray}
where
\begin{equation}
P_{\pm} =  (1 \pm \slash{v})/2 \; .
\label{proj}
\end{equation}
are projection operators.
The index $a$ runs over the light quark flavours $u, d, s$, and
the projection operators have the  property
\begin{equation}
P_{\pm}\gamma^\mu P_{\pm}  =  \pm\, P_{\pm}\, v^\mu\,P_\pm \; .  
\label{projrel}
\end{equation}

Note that in \cite{barhi,effr,itCQM,ebert2}, $G_H=1$ is used. However,
 in that case
one used a renormalization factor  for the heavy meson fields $H_v$,
which is equivalent. 

The fields $P_{5} (P_{\mu})$ annihilates a heavy-light meson,
 $0^{-}(1^-)$,  with velocity $v$.
The interaction term in the Lagrangian (equation (\ref{Int})) can, as
for the $\chi$QM, be obtained from an NJL model. In the NJL model one
starts with the Lagrangian for free quarks  and
 four quark operators, thought to be generated by gluon exchange(s).
 Taking the heavy quark limit for heavy quarks
one can obtain (after some manipulation) the interaction term (\ref{Int})
from the four quark
operator . This has been done in
\cite{effr} (-as for the light sector \cite{bijnes}).

 In our model, the hard gluons are thought to be integrated out and we are
left with soft gluonic degrees of freedom. These gluons can be
described using the external field technique, and their
effect will be parameterized by vacuum expectation values, 
i.e. the gluon condensate $\gc$. Gluon condensates with higher 
dimension could also be included, but we truncate the expansion by keeping
 only the condensate with lowest dimension.

When calculating the soft gluon effects in terms of the gluon condensate,
we follow the prescription given in \cite{nov}.
\begin{figure}[t]
\begin{center}
\epsfbox{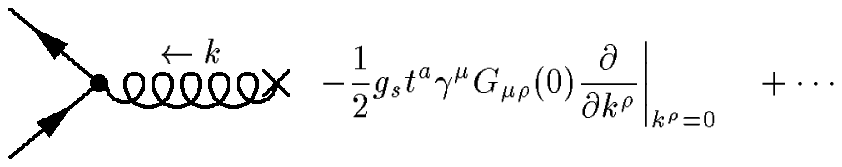}
\epsfysize=5cm
\caption{Feynmanrule for the light quark -soft gluon  vertex.}
\label{fig:gc}
\end{center}
\end{figure} 
The calculation  is easily carried out in the
Fock -  Schwinger gauge. In this gauge one can expand the gluon field as :
\begin{equation}
A_\mu^a(k) =   -  \fr{i(2\pi)^4}{2}G_{\rho\mu}(0)\fr{\partial}{\partial
k_\rho}\delta^{(4)}(k)  +  \cdots\, .
\end{equation}
Since each vertex in a Feynman diagram is accomplished with an
integration we get the Feynman rule given in figure \ref{fig:gc}.
The gluon condensate is obtained by the replacement
\begin{equation}
g_s^2 G_{\mu \nu}^a G_{\alpha \beta}^b  \; \rightarrow \fr{4 \pi^2}{(N_c^2-1)}
\delta^{a b} \gc \frac{1}{12} (g_{\mu \alpha} g_{\nu \beta} -  
g_{\mu \beta} g_{\nu \alpha} ) \, .
\end{equation}.

We observe that soft gluons
coupling to a heavy quark is suppressed by $1/m_Q$, since to leading
order the vertex is proportional to $v_\mu v_\nu G^{a\mu\nu}= 0$,
 $v_\mu$ being the heavy quark velocity.

\section{Bosonization within the HL$\chi$QM}\label{sec:strong}
 
The interaction term ${\cal{L}}_{Int}$ in (\ref{Int}) can now 
be used to bosonize the model, i.e. 
integrate out the quark fields. This can be done in the path 
integral formalism and the result is formally a functional determinant. This
determinant can be expanded in terms of Feynman diagrams, by attaching the
 external fields $H_v^{a}, \overline{H_v^{a}},  {\cal{V}}^\mu,
  {\cal{A}}^\mu$ and $\widetilde{M}_q^{V,A}$ of section II to quark loops.
Some of the loop integrals will be divergent and have to, as for the pure
 light sector \cite{chiqm,pider,epb,BEF}, to be related to physical parameters.
The strong chiral Lagrangian 
has the following form \cite{itchpt,wise1,wiseR,Grin,GriBo,IWSte,EJenk}: 
\bea
{\cal L}_{Str}\, &&= 
  \, - (1+ \frac{\varepsilon_1}{m_Q}) 
Tr\left[\overline{H_{a}}(iv\cdot D)H_{a}\right]\, + \,
 (\Delta_Q + \frac{\delta_Q}{m_Q})Tr\left[\overline{H_{a}}H_{a}\right]\,
\nonumber \\ 
 &&+\, (1+ \frac{\varepsilon_1}{m_Q}) Tr\left[\overline{H_{a}}H_{b}v_\mu {\cal V}^\mu_{ba}
\right]
 - \, (g_{\cal A} -\frac{g_1}{m_Q}) 
Tr\left[\overline{H_{a}}H_{b}\gamma_\mu\gamma_5 {\cal
A}^\mu_{ba}\right]\nonumber
\\ &&\, + 
 2 \lambda_1 Tr\left[\overline{H_{a}}H_{b} (\widetilde{M}_q^V)_{ba}\right]
+2 \lambda_1^\prime 
Tr\left[\overline{H_{a}}H_{a}\right](\widetilde{M}_q^V)_{bb}\, + .....
\nonumber \\
&& -\fr{\lambda_2}{4m_Q} 
Tr\left[\overline{H_{a}}\sigma^{\alpha\beta}H_{a}\sigma_{\alpha\beta}\right]
+\fr{\varepsilon_2}{m_Q}
Tr\left[\overline{H_{a}}\sigma^{\alpha\beta}iv\cdot DH_{a}\sigma_{\alpha\beta}\right]
 \,\nonumber \\&&
 -\, \fr{\varepsilon_2}{m_Q}
Tr\left[\overline{H_{a}}\sigma^{\alpha\beta}v_\mu {\cal V}^\mu_{ba}\sigma_{\alpha\beta}H_{b}
\right]+\, \fr{g_{2}}{m_Q}Tr\left[\overline{H_{a}}
\gamma_\mu\gamma_5 {\cal A}^\mu_{ba}H_{b}\right] \, + ....
 \label{LS1}
\eea
where the ellipses indicate other terms (of higher order, say), and 
$D_\mu$ contains the photon field.
\begin{figure}[t]
\begin{center}
   \epsfig{file=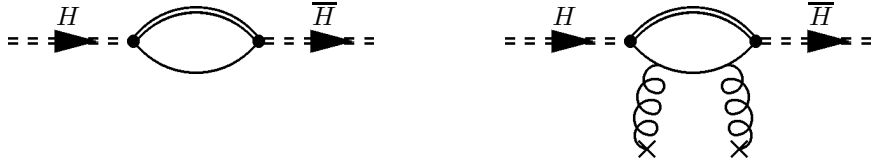}
\caption{Self energy diagrams for a heavy meson}
\label{fig:self}
\end{center}
\end{figure}
The $1/m_Q$ terms will be discarded in this section, but will be considered
later in section VI.
 Note that the term
proportional to $\Delta_Q \equiv M_H - m_Q$
 is absent in most articles considering the mesonic
 aspects only. The quantities   $\Delta_Q$,  $\lambda_1$, $\lambda_2$ and 
$\delta_Q$
 are related to the masses of  the heavy mesons. 
The trace runs over 
the gamma matrices (only). Note that in some conventions there is an extra 
 factor $M_H$ at the right- hand side in (\ref{LS1})
(our notation is the same as used in  \cite{barhi,effr,itCQM,wise1,itchpt}).

 The Feynman diagrams 
responsible for the kinetic term and the mass difference (self energy) term 
$\sim \Delta_Q$ in the meson Lagrangian is shown in figure
\ref{fig:self}. A calculation of  these two diagrams (at zero external heavy
 meson momentum), leads to the identification 
\begin{equation}
 -  iG_H^2N_c \, ( -  I_{1}  +   mI_{3/2}  +  \fr{\kappa_2}{N_c}\gc
)  -    \fr{1}{2G_3}   =    -  \Delta_Q \; ,
\label{g3rel}
\end{equation}
where we have added the last term of (\ref{Int}).
Keeping the heavy meson momentum to first order
 (subsequently to be interpreted as the derivative of the heavy meson field)
, we obtain the identification for the kinetic term:
\begin{equation}
 -  iG_H^2N_c \, (I_{3/2}  +   2mI_2  +  \fr{\kappa_1}{N_c}\gc) =   1 \; .
\label{norm}
\end{equation}

We have denoted the divergent integrals by $I$ and the finite integrals by
$\kappa$. They are defined in Appendix \ref{app:loop}. The quadratic -  ,
linear -  and logarithmic -  divergent integrals are  denoted
 $I_1$, $I_{3/2}$, and $I_2$ respectively.

The relation (\ref{norm}) is also obtained by comparison of the loop integral
 for  diagram in figure \ref{fig:va} with the vector field ${\cal V}_\mu$ 
attached to the light quark. This must  be so because of the
 relevant Ward identity, but it is also realized by explicit
 calculation.
(Note that the total covariant derivative is 
$i{{\cal D}_\mu}=iD_\mu +{{\cal V}_\mu}$  in the quark
 sector and
  $i{{\cal D}_\mu}=iD_\mu -{{\cal V}_\mu}$ in 
the meson sector
 since the meson fields transforms as an anti triplet under
 $SU(3)_V$. See Appendix \ref{app:tranf} for details)
 
From the same diagram, with the axial field ${\cal A}_\mu$ attached,
 we obtain
 the following identification for the
 axial vector coupling $g_{\cal A}$ :
\begin{figure}[t]
\begin{center}
   \epsfig{file=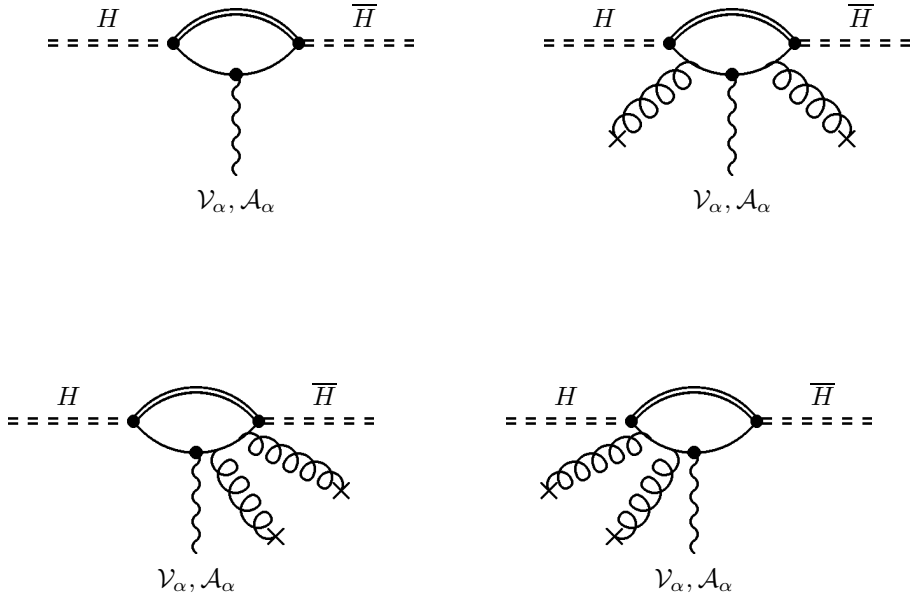}
\caption{Coupling to vector and axial vector current}
\label{fig:va}
\end{center}
\end{figure}
\begin{equation}
 g_{\cal A}\eq i G_H^2 N_c (\fr{1}{3}I_{3/2} -   2mI_2  +   \fr{4}{3}\kappa_0
  -   \fr{\kappa_1}{N_c}\gc) \; ,
\label{ga}
\end{equation}
and similarly when attaching $\widetilde{M}_q^V$
\begin{equation}
2 \lambda_1 \eq  i G_H^2 N_c (I_{3/2} -   2mI_2  + 2 \kappa_0
  -   \fr{\kappa_2}{N_c}\gc) \; ,
\label{lam1}
\end{equation}

Within the full theory (SM) at quark level, the weak current is :
\begin{equation}
J_f^\alpha  =  {\overline{q_f}_L}\,\gamma^\alpha\, Q
\label{Lcur}
\end{equation}
where $Q$ is the heavy quark field in the full theory.
Within HQEFT this current will,  below the renormalization scale
 $\mu  =   m_Q \, (=m_b, m_c)$, be modified in the following way:
\begin{equation}
J_f^\alpha
 =  {\chibar}_a \xi^{\dagger}_{af} \Gamma^\alpha  Q_v     
\, + {\cal O}(m_Q^{-1}) \; ,
\label{modcur}
\end{equation}
where \cite{neu}
\begin{eqnarray}
\Gamma^\alpha \,\eq\, C_\gamma (\mu )\,\gamma^\alpha\, L \,+\,  
C_v(\mu )\, v^\alpha\, R\; .
\label{Gamma}
\end{eqnarray}
The coefficients $C_{\gamma,v}(\mu)$ are determined 
by QCD renormalization for  $\mu < m_Q$. They have been calculated to
NLO and the result is the same in $MS$ and $\overline{MS}$ scheme\cite{Cgamma}.
Corrections to the weak current of order $1/m_Q$ will be discussed in 
section \ref{sec:mq}.

\begin{figure}[t]
\begin{center}
   \epsfig{file= 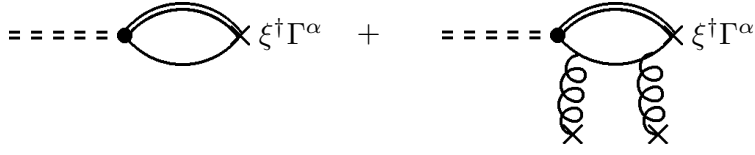,width=10cm}
\caption{Diagrams for bosonization of the left handed quark current}
\label{fig:f+}
\end{center}
\end{figure}

The operator in equation (\ref{modcur}) can be bosonized by calculating the 
Feynman diagrams shown in figure \ref{fig:f+} :
\begin{equation}
J_f^\alpha(\mbox{Bos})   =    J_f^\alpha(0) +  J_f^\alpha(1) +  \cdots
\label{Boscur}
\end{equation}

The currents $J_f^\alpha(0)$ and  $J_f^\alpha(1)$ correspond to zero
 and one axial field attached to the loop, the dots represents terms with 
more boson fields and gluon condensates. We obtain to zero order in the axial
field and first order in the gluon condensate:
\begin{equation}
J_f^\alpha (0)   =    \fr{\alpha_H}{2} Tr\left[\xi^{\dagger}_{hf}\Gamma^\alpha
  H_{vh}\right]
  \; ,\label{J(0)}
\end{equation}
where
\begin{equation}
\alpha_H \eq  -  2iG_HN_c\left(  -  I_1  +  mI_{3/2} +
 \fr{\kappa_2}{N_c}\gc\right) \; .
\label{alphaH}
\end{equation}
Note that this expression is also related to $1/G_3$ in eq. (\ref{g3rel}).
However, in the present case only $G_H$ in first power is involved.

 To next order in the chiral expansion we obtain the current
\begin{equation} 
J_f^\alpha (1) \, = \, \fr{1}{2}Tr\left[\xi^{\dagger}_{hf}\Gamma^\alpha
H_{vh}(\alpha_{H \gamma}^{(1)}\gamma^\nu\gamma_5 +
\alpha_{H v}^{(1)} v^\nu\gamma_5) {\cal A_\nu} \right] 
\label{cJ1}
\end{equation}
where the quantities $\alpha_H^{(1)}$ are given by
\bea
\alpha_{H \gamma}^{(1)}\eq \, &&\,
2iG_HN_c\left(\fr{1}{3}I_{3/2}+\fr{4}{3}\kappa_0-2mI_2+
\fr{\kappa_3}{N_c}\gc\right) \\
\alpha_{H v}^{(1)}\eq \, &&\, 2iG_HN_c\left(-\fr{2}{3}I_{3/2}
-\fr{4}{3}\kappa_0 +\fr{\kappa_5}{N_c}\gc\right) 
\eea
We observe that these quantities, as $f_H$ in (\ref{fb}), 
contains only one power of 
the coupling $G_H$, in contrast to the strong sector.

The physical meaning of the $\alpha_H$'s becomes more clear by considering their contributions to the physical quantities 
$f_H, \, f_{H^*},$  and the semileptonic form factors $f_{\pm}(q^2)$.
The coupling $\alpha_H$ in (\ref{J(0)}) is related to  the physical
 decay constants $f_H$  and $f_{H^*},$ in the following way (for $H=B,D$):
\bea
\bra 0| \overline{u}\gamma^\alpha \gamma_5 b|H \ket 
 =&   -  2\,\bra 0| J_a^\alpha|H \ket\, 
 &=  iM_Hf_H v^\alpha \; , \\
\bra 0| \overline{u}\gamma^\alpha b|H \ket 
=&   2\,\bra 0| J_a^\alpha|H \ket 
 &=  M_{H^*} f_{H^*} \varepsilon^\alpha \; .
\eea
Taking the trace over the gamma matrices in (\ref{J(0)}),
 we obtain a relation for
$\alpha_H$ and the  relations between the heavy meson decay
constants $f_H$ and $f_{H^*}$ (for  $H=B,D$) :
\begin{equation}
\alpha_H =  \fr{f_H\sqrt{M_H}}{C_\gamma (\mu ) + C_v(\mu )} =
\fr{f_{H^*}\sqrt{M_{H^*}}}{C_\gamma(\mu)}\;\;  ,
\label{fb}
\end{equation}
where the model dictates us to put $\mu  =   \Lambda_\chi$.
\begin{figure}[tb]
\begin{center}
   \epsfig{file=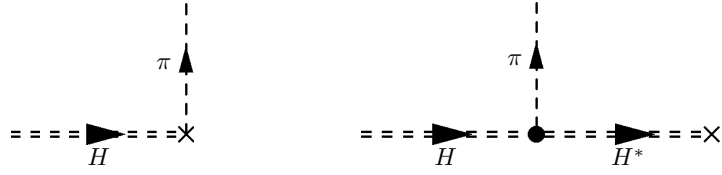,width=10cm}
\caption{Feynman diagrams for the $H\to\pi$ transitions, both
$J_a^\alpha(0)$ and $J_a^\alpha(1)$ defined in (\ref{J(0)}) and
(\ref{cJ1}) respectively, contributes to the diagram to the left } 
\label{fig:f+2}
\end{center}
\end{figure}
The  form factors  $f_+(q^2)$ and $f_-(q^2)$ are defined as :
\begin{equation}
\bra \pi^+(p_\pi )|\overline{u}\gamma^\alpha(1-\gamma_5) b|H \ket 
\, = \,  2\,\bra \pi^+(p_\pi )|J_f^{\alpha}|H \ket
\, = \, f_+(q^2)(p_H+p_\pi )^\alpha \, + \, f_-(q^2)(p_H-p_\pi )^\alpha 
\end{equation}
where $p_H^\alpha=M_H v^\alpha$ and the index $a$ corresponds to 
quark flavour $u$ and $q^\mu=p_H^\mu-k_\pi^\mu$.

The diagrams in figure \ref{fig:f+2} contributes to the $H \rightarrow
 \pi$ transition. $J_f^\alpha (0)$ and $J_f^\alpha (1)$ 
is responsible for the diagram to the left and the diagram to the
 right involve the strong Lagrangian term $\sim g_{\cal A}$ in (\ref{LS1}). 
A  calculation of the diagrams gives :
\bea
f_+(q^2) \, + \,  f_-(q^2) &=& \fr{-1}{\sqrt{2M_H}f_\pi}\left(C_\gamma +
C_v-g_{\cal A}C_\gamma \right) \alpha_H \; ,
\label{f++f-a}\\
f_+(q^2) \, -  \, f_-(q^2) &=& -C_\gamma\fr{\sqrt{M_H}}{\sqrt{2}f_\pi}
\left(\,\fr{g_{\cal A}\alpha_H}{v\cdot k_\pi} + \alpha_{H \gamma}^{(1)}\,
\right) \; ,
\label{f+-f-b}
\eea
where we have neglected terms of first order in $v \cdot k_\pi$ (where 
$\alpha_{H v}^{(1)}$ contributes). This  means that the equations are
 only valid
near the ``no-recoil point'', where $v\cdot k_\pi \rightarrow 0$ and 
$q^2 \rightarrow q^2_{\text{Max}} = (M_H^2 + m_\pi^2$).  
From equation (\ref{f++f-a}) and (\ref{f+-f-b}) we see that
$(f_+(q^2)+f_-(q^2))/(f_+(q^2)-f_-(q^2))\sim 1/M_H$, 
which is the well known Isgur-Wise scaling
law \cite{isgur2}.
 (The $1/v\cdot k_\pi$ term in (\ref{f+-f-b})
is due to the $H^*$ pole).

\section{Constraining the parameters  of the HL$\chi$QM}\label{sec:param}

Within the pure light sector the quadratic and logarithmic divergent
integrals are related to $f_\pi$ and the quark condensate in the following way
\cite{pider,epb,BEF,ebert3}:
\begin{equation}\label{I2}
f_\pi^2 =    -  i4m^2N_cI_2 +  \fr{1}{24m^2}\gc \; ,
\end{equation}
\begin{equation}\label{I1}
\qc = -4imN_cI_1-\fr{1}{12m}\gc \; ,
\end{equation}
which are obtained by relating loop diagrams to physical quantities
as for the eqs. (\ref{norm}) and (\ref{ga}). (Here the a priori 
divergent integrals $I_{1,2}$ have to be interpreted as  the regularized ones) 
As the pure light sector is a part of our model, we have to keep these 
relations in the heavy light case studied here.
 In addition, in
 the heavy-light sector the linearly divergent integral $I_{3/2}$ 
will also appear. 
As $I_1$ and $I_2$ are related to the quark
condensate and $f_\pi$ respectively, the (formally) 
linearly divergent integral  
$I_{3/2}$ is related to
$\delta g_{\cal A} \equiv 1  -  g_{\cal A}$, which is found
 by eliminating $I_2$
from eqs. (\ref{norm}) and (\ref{ga}) :
\begin{equation}
 \delta g_{\cal A} =   - \frac{4}{3} i G_H^2N_c \, \left(I_{3/2} 
 +   \kappa_0 \right)  \; .
\label{dga}
\end{equation}
Note that the gluon condensate drops out here.
Within a primitive cut-off regularization, $I_{3/2}$ is (in the leading
 approximation) proportional to the cut-off in first power \cite{barhi}.
 Within  dimensional regularization it is finite. We will keep $I_{3/2}$
as a free parameter to be determined by the physical value of $g_{\cal A}$.

Eliminating $I_{3/2}$ from the 
eqs. (\ref{norm}) and (\ref{ga}) 
and inserting the expression for $I_2$ obtained from (\ref{I2}) 
we find  the following expression for
$G_H$ :  
\begin{equation}
G_H^2 =  \fr{m(1 +  3g_{\cal A})}{2f_\pi^2 +   \fr{m^2N_c}{4\pi} -  
\frac{\eta_1}{m^2}\gc}\, ,
\quad\text{where}\quad
\eta_1\eq \fr{\pi}{32} \, .
\label{gh}
\end{equation}
Note that $G_H$ has dimension $(mass)^{- 1/2}$.

In order  to constrain the parameters further, we will consider the
 parameters  $\lambda_1$,
$\lambda_2$ and $\delta_Q$ related to the meson masses.
The gluon condensate can be related to the chromomagnetic
interaction : 
\begin{equation}
\mu_G^2(H) =  \fr{1}{2M_H}C_{\text{M}}(\mu)
\bra H|\bar{Q_v}\fr{1}{2}\sigma\cdot GQ_v|H\ket\, , \label{gc2}
\end{equation} 
where the coefficient $C_{\text{M}}(\mu)$ contains the short distance
 effects down to
the scale $\mu$ and has been calculated to next to leading order ($NLO$) 
\cite{neub,grozin1}, and can be found in table \ref{tab:predicted} 
($2 M_H$ is a normalization factor).
The chromomagnetic operator is responsible for the splitting between
the $1^- $ and $0^-$ state, and is  known from spectroscopy, 
\bea
\mu_G^2(H)  =  \,  3 \lambda_2  \, = \, \frac{3}{2} m_Q (M_{H^*} -  M_H)
 \; .  
\label{mug2exp}
\eea 

An explicit calculation of the matrix element 
 in equation (\ref{gc2}) gives
\begin{equation}
\label{mug2}
\mu_G^2  =   \eta_2 \frac{G_H^2}{m} \gc \; , \quad 
\text{where}\quad\, \eta_2\, \eq\, 
\fr{(\pi +  2)}{32}C_M(\Lambda_\chi)\, .
\end{equation}

Combining  eq. (\ref{gh}) and eq. (\ref{mug2}) we get the following relations :
\begin{equation}
\label{gcgh}
\gc  =   \fr{\mu_G^2 f_\pi^2}{2\eta_2} \, \fr{1}{\rho}  \; \, , \qquad  
G_H^2  =   \fr{2m}{f_\pi^2} \, \rho \; \, , 
\end{equation}
where the quantity $\rho$ is of order one and  given by
\begin{equation}
\label{rho}
\rho \, \eq \, \fr{(1 +  3g_{\cal A}) + 
 \frac{\eta_1 \mu_G^2}{\eta_2 m^2}}{4 (1 + \frac{N_c m^2}{8 \pi f_\pi^2} )}
\; \, .
\end{equation}

In the limit where only the leading logarithmic integral $I_2$ is kept
in (\ref{norm}),
 we  obtain:  
\begin{equation}
g_{\cal A} \rightarrow \, 1 \; , \qquad
\rho \rightarrow \, 1 \; , \qquad  
G_H \; \rightarrow \; G_H^{(0)} \, \eq \, \fr{\sqrt{2m}}{f_\pi} \; \, . 
\label{IRlim}
\end{equation}
Note that $g_{\cal A}= 1$ is the non-relativistic value \cite{itchpt}.

The quantity $\delta_Q$
 is found  from the kinetic term  as:
\begin{equation}
\delta_Q \,= \, \mu_\pi^2 \eq \fr{1}{2M_H}\bra H|\overline{Q_{v}} 
(D_\perp)_{\text{eff}}^2Q_{v}|H\ket\,
\, , \; \text{where} \quad (D_\perp)_{\text{eff}}^2 \eq D^2- C_K (v\cdot D)^2
\end{equation}

This quantity can easily be calculated in our model, and 
a direct calculation gives the expression:
\bea
\mu_\pi^2&\,=\,&iG_H^2N_cm^2\left\{
I_{3/2}+2mI_2+\fr{\kappa_1}{N_c}\gc\right\}
+ \eta_3 \fr{G_H^2}{m}\gc\nonumber \\ &&+   
\fr{1}{4} C_K G_H^2 \left\{-4imN_cI_1-\fr{1}{12}\gc\right\} \, , \;
\text{where} \;
\eta_3 = \frac{5}{48}+\frac{\pi}{64} \; .
\label{mupi1}
\eea

Note that the $(v \cdot \partial)^2$ part of the kinetic term cancels the
 two heavy quark propagators, and the light quark condensate (\ref{I1})
 appears naturally.
Eliminating the divergent integrals, using equation (\ref{norm}),
 (\ref{I1}) and (\ref{mug2}), some of the $G_H$'s can be included in
 physical parameters, and  we obtain  
\begin{equation}
\delta_Q = \mu_\pi^2 =  \fr{\eta_3}{\eta_2} \mu_G^2  \, - m^2
 - \fr{1}{4} C_K \qc G_H^2 \, .
\label{mupi2}
\end{equation}

Note that the last term (originating from $(v\cdot D)^2Q_v$ at quark level)
gives a vanishing contribution  in the free quark limit when
 $G_H \rightarrow 0$.
Using the expressions for $\gc$ and $G_H^2$ in (\ref{gcgh}),
 we find an expression for $g_{\cal A}$ in terms of $m$,
$\mu_\pi^2$ and $\qc$.
However, as $\mu_\pi^2$
is not very well known, and $C_K$ is known only to leading logarithmic 
approximation, we will not try to determine  $g_A$ by this relation.

From eqs (\ref{lam1}), (\ref{I2}) and (\ref{mug2}), we find
\begin{equation}
\label{lamb-ga}
2\lambda_1 = \frac{1}{2}(3 g_{\cal A} -1) - 
\fr{(9 \pi -16)  \mu_G^2}{384\eta_2 m^2} \, .
\end{equation}
 In the limit (\ref{IRlim}) we obtain 
$2 \lambda_1 \rightarrow 1$, as expected. The parameter $\lambda_1$
is related to the mass difference $M_{H_s}- M_{H_d}$. Unfortunately, this 
cannot be used at the present stage to constrain the parameters  in our 
model because  this quantity has large chiral corrections \cite{EJenk}.

Using equation (\ref{norm}) and (\ref{I1}) we can write $\alpha_H$ as:
\begin{equation}\label{qcrel}
\alpha_H=\fr{G_H}{2}\left(-\fr{\qc}{m}-2f_\pi^2(1-\fr{1}{\rho})+
\fr{(\pi-2)}{16m^2}\gc\right) \; ,
\end{equation}
or equivalently, using the eqs. (\ref{I1}), (\ref{dga}) and (\ref{gc2}) as
\begin{equation}\label{qcrel2}
G_H \alpha_H= \fr{3}{2} \, m (1- g_{\cal A}) + 
\fr{(3 \pi +4)}{192 \, \eta_2} \,  \fr{\mu_G^2}{m} + 
G_H^2 \left(-\fr{\qc}{2 m}+ \fr{m^2}{8 \pi} \right) \; .
\end{equation}
Note that the relation (\ref{fb}) gives $f_{H} >  f_{H^*}$, which is not 
correct experimentally. Adding  $1/m_Q$  corrections we correctly  
reproduce   $f_{H^*} >  f_{H}$ \cite{GriBo}.

Combining (\ref{fb}) with (\ref{qcrel}), we obtain \cite{EFZ} in the
 leading limit (taking into account the logarithmic and quadratic divergent
  integrals only, and 
let $C_\gamma \rightarrow 1$, $C_v \rightarrow 0$ and 
$g_{\cal A} \rightarrow 1 $ as in (\ref{IRlim}) ) : 
\begin{equation}
 f_H\sqrt{M_H}\; \rightarrow \,  -  \frac{\qc}{f_\pi \sqrt{2m}} \;\;  ,
\label{GTrel}
\end{equation}
which gives the  scale for $f_H$ (It is, however, numerically a factor
2 off for the $B$-meson).

Using the relations in equation (\ref{norm}), (\ref{ga}) and
(\ref{I2}) we obtain for  $\alpha_{H \gamma}^{(1)}$ and $\alpha_{H v}^{(1)}$ :
\bea
\alpha_{H \gamma}^{(1)}&&=\fr{2 g_{\cal A}}{G_H}\label{alpha1} \; , \\
\alpha_{H v}^{(1)}&&=\fr{4}{3} G_H 
\left(\fr{f_\pi^2}{2m} \left(\fr{1}{\rho}-1 \right)
+\left[\fr{mN_c}{8\pi}+\fr{(\pi+8)}{256m^3}\gc \right] \right) \; ,
\eea
where the latter may also be written: 
\bea
G_H \alpha_{H v}^{(1)} = (1 - g_{\cal A}) - \fr{1}{3} G_H^2
\fr{mN_c}{4\pi} + \fr{(\pi+8) \mu_G^2}{256 \eta_2 m^2} \; ,
\eea

Combining equation (\ref{f++f-a}) and
(\ref{f+-f-b}), we find (up to terms of first order in $v \cdot k_\pi$):
\bea
f_+(q^2)&=&
-\fr{g_{\cal A}f_{H^*}\sqrt{M_H M_{H^*}}}{2\sqrt{2}f_\pi v\cdot k_\pi}-
C_\gamma\fr{ g_{\cal A}\sqrt{M_H}}{G_H\sqrt{2}f_\pi}
-\fr{1}{2\sqrt{2}f_\pi}\left\{f_H-g_{\cal A}f_{H^*}
\sqrt{\fr{M_{H^*}}{M_H}}\right\} \; ,
\label{f+}
\eea
where we have used equation (\ref{fb}) and (\ref{alpha1}).

\section{The Isgur-Wise function}\label{sec:iw}

\begin{figure}[t]
\begin{center}
\epsfig{file=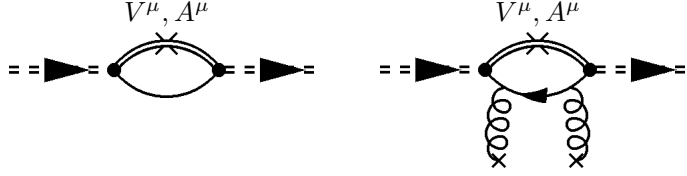,width=10cm}
\caption{Loop diagrams for bosonizing the $b \rightarrow c$ current,
$V^\mu=\gamma^\mu$, $A^\mu=\gamma^\mu\gamma_5$}
\label{fig:iw}
\end{center}
\end{figure}

The Isgur-Wise function \cite{isgur},
 $\xi (\omega)$, relates all the
form factors describing the processes $B\to D(D^*)$ in the
heavy quark limit :
\bea
\fr{\bra D|\overline{Q_{cv}}\gamma^\mu Q_{bv}|B\ket}{\sqrt{M_BM_D}}
&&=\xi (\omega )(v^\mu+{v^\prime}^\mu) \, , \; \omega \, \eq \, v \cdot
v^\prime \; , \nonumber \\
\fr{\bra D^*|\overline{Q_{cv}}\gamma^\mu\gamma_5 Q_{bv}|B\ket}{\sqrt{M_BM_{D^*}}}
&&=\xi(\omega)\left[ {v^\prime}^\mu\varepsilon^*\cdot v 
-{\varepsilon^*}^\mu(1+\omega)\right]  \; , \\
\fr{\bra D^*|\overline{Q_{cv}}\gamma^\mu Q_{bv}|B\ket}{\sqrt{M_BM_{D^*}}}
&&=\xi(\omega)i\varepsilon^{\mu\nu\lambda\sigma}{\varepsilon^*}_\nu
v^\prime_\lambda v_\sigma \; ,
\eea
where $Q_{cv}$ and $Q_{bv}$ are the $c$ and $b$ quark fields within
HQEFT.
The Isgur-Wise function (IW) can be calculated straight forward
 by calculating the
diagrams shown in figure (\ref{fig:iw}). The result is 
\begin{equation}\label{iw}
\xi(\omega)= \fr{2}{1+\omega} \left(1-\rho \right)+
 \rho \, r(\omega)\, , 
\end{equation}
where $\rho$ is given in (\ref{rho}) and 
$r(\omega)$ is the same function appearing in loop calculations of the
anomalous dimension in HQEFT :
\begin{equation}
r(\omega)=\fr{1}{\sqrt{\omega^2-1}}\, 
\text{ln}\left(\omega+\sqrt{\omega^2-1}\right)
\end{equation}
We see that $\xi$ is normalized to 1 at zero recoil. Note that
equation (\ref{iw}) is only valid in the limit of equal gluon
condensate and the coupling $G_H$ in the $B$ and $D$ sector. The derivative of
the IW function at zero recoil is :
\begin{equation}
\left.\fr{\partial \xi}{\partial
w}\right|_{w=1}=-\fr{1}{2}+\fr{1}{6} \, \rho
\end{equation}

In the limit (\ref{IRlim}), the
IW function takes the simple form :
\begin{equation}\label{iwsimp}
\xi(\omega)=r(\omega)\qquad
\text{and}\qquad
\left.\fr{\partial \xi}{\partial
w}\right|_{w=1}=-\fr{1}{3}
\end{equation}
Numerically, the full result (\ref{iw}) is only a few percent away
from (\ref{iwsimp}). 
Adding the short distance QCD effects to (\ref{iw}) will slightly  modify  our
result \cite{neub}. For the IW function describing $B\to
D^*$ transition, the derivative at $\omega =1$ gets a contribution
 $-0.07$ from QCD corrections such that $\xi^\prime(0)\simeq -0.40$.

\section{$1/m_Q$ corrections within the HL$\chi$QM}\label{sec:mq}

\subsection{Bosonization of the strong sector}

\begin{figure}[t]
\begin{center}
   \epsfig{file=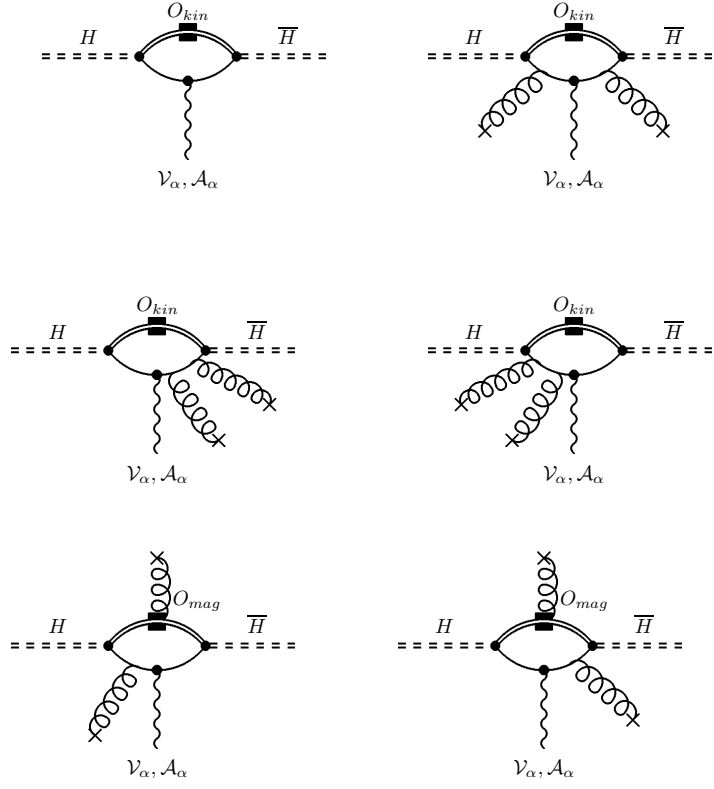,width=9cm}
\caption{Diagrams responsible for $1/m_Q$ terms in the chiral Lagrangian}
\label{fig:kin}
\end{center}
\end{figure}

The new terms of order $1/m_Q$ in (\ref{LS1})are a consequence of
 the chromomagnetic interaction
(the second term in equation (\ref{LHQEFT})), which break heavy quark
spin symmetry, and the kinetic term (the third term in (\ref{LHQEFT})).
 Calculating the diagrams of figure \ref{fig:kin} gives the 
 following identifications :
\bea
\varepsilon_1&=&-  iG_H^2N_c \left(I_1-mI_{3/2}+2m^2I_2-m\kappa_0 
\right.\\ \nonumber &&\qquad\qquad\left. +
\fr{C_K}{4}(I_1-m^2I_2)+\fr{\kappa_4}{N_c}\gc \right) \label{corrt1} \\
g_1&=& -iG_H^2N_c\left(-\fr{1}{3}I_1+mI_{3/2}
+\fr{4}{3}m^2I_2+m\kappa_0-\fr{2}{3} \widetilde{\kappa_0} 
\right.\\ \nonumber &&\qquad\qquad\left.
+\fr{C_K}{4}(I_1+3m^2I_2)+\fr{\kappa_5}{N_c}(2 +C_K) \gc \right)
\label{corrt2} \\
g_2&=& C_M(\Lambda_\chi)\fr{(\pi+4)}{192m^2}G_H^2\gc \qquad ; \qquad
\varepsilon_2=-\fr{g_2}{2} \; . 
\eea

As the $1/m_Q$ terms break heavy quark spin symmetry, the chiral Lagrangian
in (\ref{LS1}) will split in $H(0^-)$ and $H^*(1^-)$ terms respectively.
 However, 
by allowing for a flavour and spin dependent renormalization constants we can
write the Lagrangian  in the compact form: 
\begin{equation}
{\cal L} =  -Tr\left[\overline{H^{\text{r}}_{a}}(iv\cdot {\cal D}_{ba}
 -\Delta_Q)^{\text{r}} H^{\text{r}}_{b}\right]\, -\, \gat 
Tr\left[\overline{H^{\text{r}}_{a}}H^{\text{r}}_{b}
\gamma_\mu\gamma_5 {\cal A}^\mu_{ba}\right]\, ,
\end{equation}
where 
$i{\cal D}^\mu_{ba}=i \delta_{ba} D^\mu-{\cal V}^\mu_{ba}$ and we have
 redefined the $H$ fields as 
$H=H^{\text{r}}\sqrt{Z_H}$, where $Z_H$ and the new coupling $\gat$ are
now defined :
\bea
Z_H^{-1}  &=& 1+\fr{\varepsilon_1-2d_M\varepsilon_2}{m_Q} \qquad , \;
\Delta_Q^{\text{r}} = Z_H \left( \Delta_Q + \fr{\mu_\pi^2}{m_Q} \right) \\
\gat &=& \ga\left(1-\fr{1}{m_Q}(\varepsilon_1-2d_{\cal A}\varepsilon_2)\right)
-\fr{1}{m_Q}(g_1-d_{\cal A}g_2)
\eea
where
\begin{equation}
d_M=\begin{cases}\,\,\,\,\, 3\quad\text{for}\quad 0^-\\
                 -1\quad\text{for}\quad 1^-
\end{cases}\quad\quad
d_{\cal A}=\begin{cases}\,\,\,\, 1\quad\text{for}\quad H^*H\,\,\,\quad\text{coupling}\\
              -1\quad\text{for}\quad H^*H^*\quad\text{coupling}
\end{cases}
\end{equation}
The last term in equation (\ref{LS1}) gives a splitting in the mass
of the $H^*(1^-)$ and $H(0^-)$ state.  This term can be absorbed in a
redefinition of $\Delta_Q$, namely  $\Delta_Q \to \Delta_Q-d_M\Delta_H/4$, 
where $\Delta_H \eq M_{H^*}-M_H$ (Note that $\mu_G^2=3m_Q\Delta_H/2$).

Eliminating the divergent integrals in (\ref{corrt1}) and (\ref{corrt2})
using equations  (\ref{I2}) (\ref{I1}) and (\ref{dga}),
we can rewrite $\varepsilon_1$ , $g_1$ and  $g_2$ as :
\bea
\varepsilon_1&=&-m+G_H^2\left(\fr{\qc}{4m}+f_\pi^2+\fr{N_cm^2}{16\pi}+\nonumber
\fr{C_K}{16}(\fr{\qc}{m}-f_\pi^2)\right.\\&&\left.\qquad\qquad\quad
+\fr{\eta_5}{m^2}\gc\right)\quad\text{where}\quad\eta_5\eq\fr{1}{128}(C_K+8-3\pi)\\
g_1&=&\,m-G_H^2\left(\fr{\qc}{12m}+\fr{f_\pi^2}{6}+
\fr{N_cm^2(3\pi+4)}{48\pi}- \nonumber
\fr{C_K}{16}(\fr{\qc}{m}+3f_\pi^2)\right.\\&&\left.\qquad\qquad\quad
+\fr{\eta_6}{m^2}\gc\right) \; ,\quad\text{where}\quad
\eta_6\eq\fr{1}{64}(C_K-2\pi)  \; , \\
g_2&=& \fr{(\pi + 4)}{(\pi+2)} \fr{\mu_G^2}{6 m}
\; .
\eea
The masses of the particles now before SU(3) breaking terms are 
taken into account
\bea
\label{mesmas}
M(0^-) = m_Q + \Delta_Q + \fr{(\mu_\pi^2 - \mu_G^2)}{m_Q} \\
M(1^-) = m_Q + \Delta_Q + \fr{(3 \mu_\pi^2 + \mu_G^2)}{3 m_Q} \; . 
\eea

\subsection{Bosonization of the weak current}\label{sec:mqweak}

In HQEFT the weak vector current at order $1/m_Q$ is \cite{neu}:
\begin{equation}
J_V^\alpha=\sum_{i=1,2}C_i(\mu)J_i^\alpha+\fr{1}{2m_Q}\sum_jB_j(\mu)O_j^\alpha
+\fr{1}{2m_Q}\sum_kA_k(\mu)T_k^\alpha \; ,
\end{equation}
where the first terms are given in (\ref{modcur}) and (\ref{Gamma}), 
the $B_j$'s and $A_j$'s are Wilson coefficients, and the $O_j^\alpha$'s
are two quark operators
\bea
& O_1^\alpha=\bar{q}_L\gamma^\alpha i\slash{D}Q_v \; , \quad & 
O_4^\alpha=\bar{q}_L\gamma^\alpha 
(-iv\cdot\overleftarrow{D})Q_v \; , \nonumber\\
& O_2^\alpha=\bar{q}_Lv^\alpha i\slash{D}Q_v\quad \; , 
& O_5^\alpha=\bar{q}_Lv^\alpha 
(-iv\cdot\overleftarrow{D})Q_v\nonumber \; , \\
& O_3^\alpha=\bar{q}_LiD^\alpha Q_v \; , \quad & O_6^\alpha=\bar{q}_L 
(-i\overleftarrow{D}^\alpha)Q_v   \; ,
\label{B}
\eea
The operators $T_k$ are nonlocal and is
a combination of the leading order currents $J_i$ and a term of order
$1/m_Q$ from the effective Lagrangian (\ref{LHQEFT}):
\bea
\fr{T_1^\alpha}{2m_Q}&=&i\int dy^4 
T\{J_1^\alpha(0),O_{\text{kin}}(y)\} \; ,\nonumber \\ 
\fr{T_2^\alpha}{2m_Q}&=&i\int d^4y 
T\{J_2^\alpha(0),O_{\text{kin}}(y)\} \; ,\nonumber \\
\fr{T_3^\alpha}{2m_Q}&=&i\int d^4y 
T\{J_1^\alpha(0),O_{\text{mag}}(y)\} \; , \nonumber \\ 
\fr{T_4^\alpha}{2m_Q}&=&i\int d^4yT\{J_2^\alpha(0),O_{\text{mag}}(y)\} \; ,
\eea
where 
\begin{equation}
O_{\text{kin}}\,\eq\,\fr{1}{2m_Q}\bar{Q}_v\,(iD_\perp)_{\text{eff}}^2\,Q_v
 \; ,\qquad \text{and}\qquad
O_{\text{mag}}\,\eq\,-\fr{g_s}{4m_Q}\bar{Q}_v\,\sigma\cdot G\, Q_v \; .
\end{equation}
Bosonizing the $O_j^\alpha$ operators in (\ref{B}) we obtain:
\begin{equation}
 \fr{1}{2}Tr\{\tilde{\Gamma}^{\alpha\mu} H\xi^\dagger
(\alpha_3^\gamma\gamma_\mu+\alpha_3^vv_\mu)\}
\end{equation}
where $\tilde{\Gamma}^{\alpha\mu}$ is defined :
\bea
&\tilde{\Gamma}^{\alpha\mu}&=B_1\,\gamma^\alpha
\,L\,\gamma^\mu\, + \,B_2\,v^\alpha \,R\,\gamma^\mu \, + \,
B_3\,g^{\alpha\mu}\,R\,\,\nonumber \\ &&
 + \,B_4\,\gamma^\alpha\, L\,v^\mu\, + \,B_5\,v^\alpha\, R\,v^\mu \, + \, B_6\,g^{\alpha\mu}\,R
\eea
The coefficients $\alpha_3^\gamma$ and $\alpha_3^v$ can be written
after the use of   
(\ref{I2}) (\ref{I1}) and (\ref{dga}) :
\bea
&\alpha_3^\gamma&=2G_H\fr{m^2}{3}\left\{\fr{f_\pi^2}{2m}(\fr{1}{\rho}-1)+
\fr{\pi- 2}{64m^3}\gc\right\}
=\fr{m}{3}\alpha_H+\fr{G_H}{6}\qc\label{alphag3}\\
&\alpha_3^v&=\alpha_3^\gamma+\fr{\qc}{2}G_H
=\fr{m}{3}\alpha_H+\fr{2}{3}G_H\qc\label{alphav3}
\eea

Bosonization of these non-local operators is straightforward within our
 model and the result is 
\bea
&& \bra 0| (A_1T_1^\alpha + A_2T_2^\alpha) |H \ket  = -\fr{\mu_\pi^2}{G_H}
Tr\{\xi^\dagger\Gamma^\alpha H_v\}\\
&& \bra 0| (A_3T_3^\alpha + A_4T_4^\alpha) | H \ket = \fr{\mu_G^2d_M}{3G_H}
Tr\{\xi^\dagger\Gamma^\alpha H_v\}\, ,
\eea
where $\Gamma^\alpha$ is given in (\ref{Gamma}).

When $1/m_Q$ terms are included, we find that (\ref{fb}) is modified 
in the following way:
\bea
f_H\sqrt{M_H}&=&\alpha_H(C_\gamma+C_v)
(1-\fr{\e_1-6\e_2}{2m_Q})\nonumber \\ 
&&+\fr{B_\gamma\alpha_3^\gamma+B_v\alpha_3^v}{2m_Q}
-(C_\gamma+C_v)\fr{(\mu_\pi^2-\mu_G^2)}{G_H m_Q}\\
f_{H^*}\sqrt{M_{H^*}}&=&\alpha_HC_\gamma
(1-\fr{\e_1+2\e_2}{2m_Q})\nonumber
\\ &&+\fr{B_\gamma^*\alpha_3^\gamma+B_v^*\alpha_3^v}{2m_Q}
-C_\gamma\fr{(3\mu_\pi^2+\mu_G^2)}{3G_H m_Q}
\eea
where
\bea
&&B_\gamma\,\,\eq\, 2B_1-4B_2-B_3+B_4-B_5-B_6\; ,\nonumber \\
&&B_v\,\,\eq\, B_1+B_2+B_3+B_4+B_5+B_6 \nonumber	\; ,\\
&&B^*_\gamma\,\eq\, -2B_1+B_3-B_4+B_6 \qquad\text{and}\qquad
B^*_v\,\eq\, B_1+B_4 \; .
\eea
The ratio between the coupling constants is :
\bea
\fr{f_{H^*}}{f_H}&=&\fr{C_\gamma}{C_\gamma+C_v}
\left\{1-\fr{1}{m_Q}\left(4\e_2+\fr{4\mu_G^2}{3\alpha_HG_H}\right)\right\}
\nonumber\\ &&+\fr{1}{2m_Q\alpha_H}\left((B^*_\gamma-B_\gamma)\alpha_3^\gamma
+(B^*_v-B_v)\alpha_3^v\right) \; .
\eea

\section{Chiral corrections}\label{sec:chiral}

Chiral corrections are numerically comparable with the $1/m_Q$ corrections.
While the relevant mass scale of $1/m_Q$ corrections are 
$\Delta_H=M_{H^*}-M_H$ (cfr. eq (\ref{mesmas})),
 the relevant mass scale of chiral corrections are 
$\delta=M_{H_s}-M_{H_{u,d}}$.
Clearly  $\Delta_H \sim \delta$ numerically.
 In this section we will  consider just the
chiral corrections to $f_H$ and $\ga^{H^*H\pi}$ which are necessary to include
in the numerical analysis. These corrections
have been calculated by many groups \cite{GriBo,goity,cheng2}
, the result is in $\overline{MS}$ scheme, including counter terms :
\bea
&&f_{H_{u,d}}^{\chi}\sqrt{M_{H_{u,d}}}=
f_{H_{u,d}^*}^{\chi}\sqrt{M_{H^*_{u,d}}}=
\nonumber\\
&&\alpha_H \left(
1-\fr{1}{32\pi^2f^2}\fr{11}{18}\left\{
-m_K^2(1+\ga^2) 
+m_K^2(\ln m_K^2/\mu^2+\fr{2}{11}\ln\fr{4}{3})(1+3\ga^2) \right\}\right)\; , \\
&&f_{H_{s}}^{\chi}\sqrt{M_{H_{s}}}=f_{H_{s}^*}^{\chi}\sqrt{M_{H^*_{u,d}}}=
\nonumber\\
&&\alpha_H\left(
1-\fr{1}{32\pi^2f^2}\fr{13}{9}\left\{-m_K^2(1+\ga^2)
+m_K^2(\ln m_K^2/\mu^2+\fr{4}{13}\ln\fr{4}{3})(1+3\ga^2) \right\}
\nonumber\right.\\
 &&\left.
\qquad\quad-\fr{\omega_1f^2}{\alpha_H\qc}m_K^2
\right) \; .
\eea
Note that this result is independent of $\Delta_Q$.
We have ignored terms proportional to $m_\pi^2$ and used the mass relation
$m_{\eta_8}^2=4 m_K^2/3$. The counter terms needed to make the
expression above finite originates from
the following terms in the weak current :
\begin{equation}
J^\mu_a({\cal M}) = 
\fr{\omega_1}{2}Tr[\xi^\dagger_{ba}\,\Gamma^\mu {H_v}_{c}\,
\widetilde{{\cal M}^V_{cb}}]
+\fr{\omega_1^\prime}{2}Tr[\xi^\dagger_{ba}\,\Gamma^\mu {H_v}_{b}]\,
\widetilde{{\cal M}^V_{cc}} \; ,
\end{equation}
where $\Gamma^\mu$ is given in (\ref{Gamma}). Moreover,
 $\omega_1=-4\lambda_1/G_H$, where $\lambda_1$ is given in equation
(\ref{lam1}). We ignore  $\omega_1^\prime$ since it is subleading in $1/N_c$.
(This is similar to what is found for $2L_1-L_2$, $L_4$ and $L_6$ in
the light sector \cite{chiqm}). Therefore, in the limit where we neglect $u,d$
quark masses $f^\chi_{H_{u,d}}$ does not depend on counterterms. This
also happens for $\ga^{H^*H\pi}$. The chiral corrections to $\ga$ can
be calculated with the formula listed in appendix \ref{app:loop} :
\bea
g^\chi_{\cal A}=&&\ga\left\{1-\ga^2\fr{1}{32\pi^2f^2}\left(\fr{35}{9}m_K^2\ln
\fr{m_K^2}{\mu^2}+\fr{8}{9}m_K^2\ln\fr{4}{3}-\fr{5}{3}m_K^2
\right.\right.\nonumber\\&&\left.\left.
\qquad-
\fr{\Delta_Q^2}{3}\left(\ln\fr{4m_K^2}{3\mu^2}-2F(m_{\eta_8}/\Delta_Q)
\right)\right.)\right\} \; .
\eea
$F(x)$ is defined in (\ref{F}). This result coincides with the one in
\cite{cheng2} for $\Delta_Q=0$. It turns out that for $\Delta_Q\sim
(0-0.5)$ GeV, the chiral corrections vary with less than $1\%$. Therefore 
 we will simply ignore $\Delta_Q$.

Note that in the expressions for the chiral corrections to $f_H$ and $g_A$
considered in this section, the $1/m_Q$ corrections of the preceding section
are not included. However, both chiral and $1/m_Q$ corrections are of
 course included in our numerical analysis in the next section.

\section{Numerical Results}\label{sec:num}

 As we have seen in the  previous sections, our bosonizing procedure
 puts restrictions, in the form of relations between the the model dependent
parameters 
 $m$, $G_H$, $\gc$, $\qc$  and the measurable 
parameters (quantities) such as $f_\pi$, $f_H$,
 $g_{\cal A}$, $\mu_G^2$, 
and $\mu_\pi^2$. We will use a
 standard value of the quark condensate, $\qc = -(0.240\,
 GeV)^3$. (It is not clear if the quark and the gluon condensates 
defined in this paper
are  the same as those in QCD sume rules).
 In principle, we have enough relations to
fix the model dependent parameters. However, some physical quantities are
 still relatively uncertain, which means that the values of the model
 dependent parameters cannot be given a very precise value.

In principle, $\Delta_Q = M_H - m_Q$ is also a parameter of the model,
which enters if we calculate diagrams with external momenta ( $v^\mu \Delta_Q$
will then be a part of an external momentum). However, the way we are 
bosonizing here, the external fields (${\cal V}, {\cal A}$ and $H_v$)
carry zero external momenta. Then
 $\Delta_Q$ will not enter our loop integrals, 
and eq. (\ref{g3rel}) is irrelevant (so far) within our model. In the
case of chiral corrections $\Delta_Q$ could play a role. However, as we
have seen in the previous section, $\Delta_Q$ does not enter in the
chiral corrections to $f_H$ and plays a very little numerical role for
the corrections to $\ga$.

In the chiral corrections to $f_H$ and $\ga$, we have consequently
used $\overline{MS}$ as in \cite{BEF}. In pure $\chi$PT
the ``$(\overline{MS} +1)$'' scheme is used.
We have 
explicitly checked that the numbers in table \ref{tab:predictedb} and 
\ref{tab:predictedd} can be reproduced in ``$(\overline{MS}+1)$'' with a small
change in $m$ and $\ga$ (the bare coupling constant).

\begin{figure}[t]
\begin{center}
   \epsfig{file=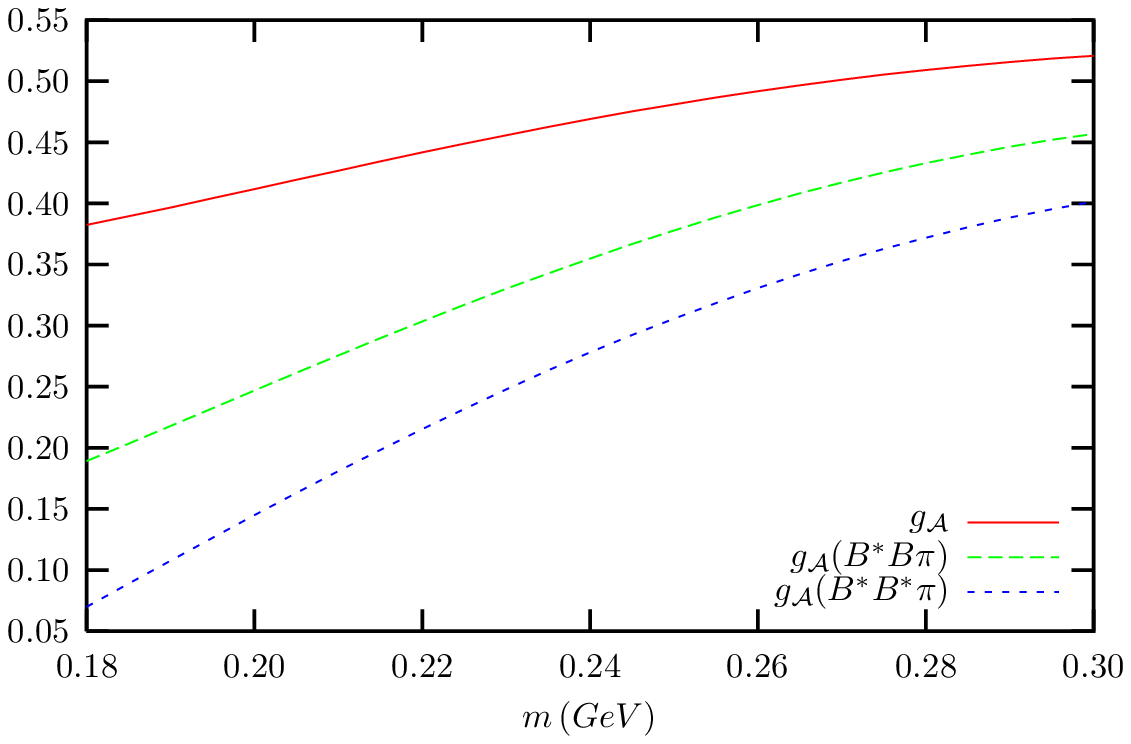}
\caption{The strong coupling constant in the $B$-sector}
\label{fig:Ga}
\end{center}
\end{figure}

\begin{figure}[t]
\begin{center}
   \epsfig{file=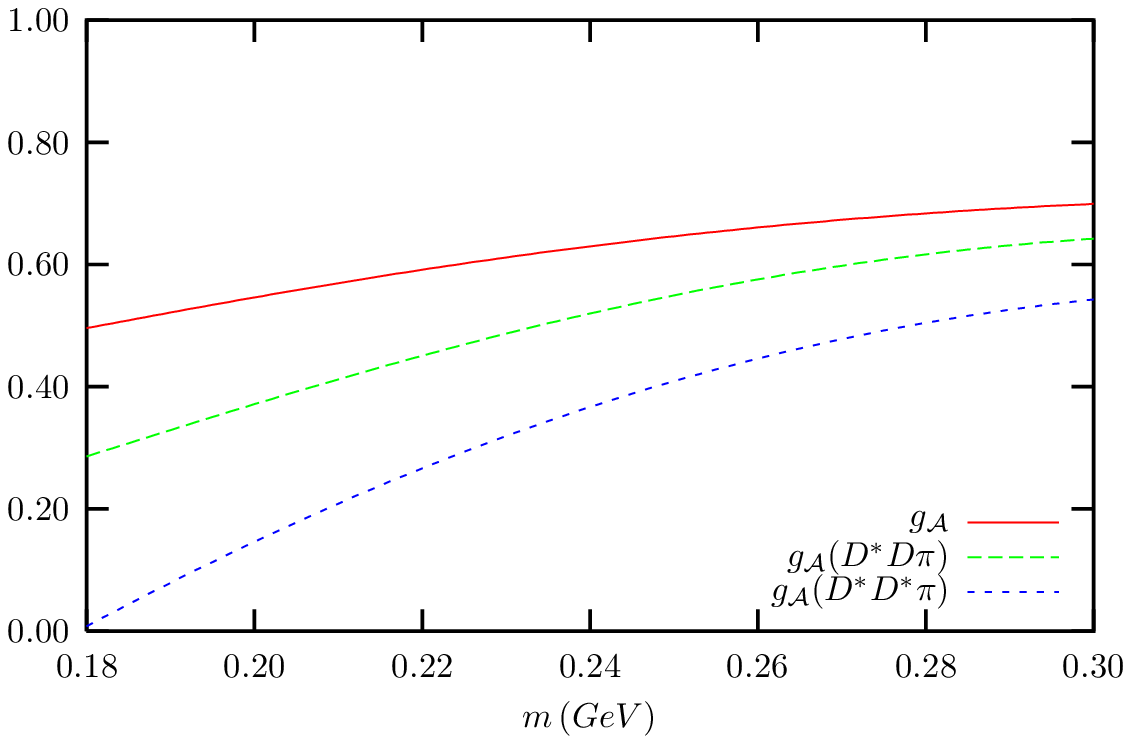}
\caption{The strong coupling constant in the $D$-sector}
\label{fig:Gac}
\end{center}
\end{figure}

\begin{figure}[t]
\begin{center}
   \epsfig{file=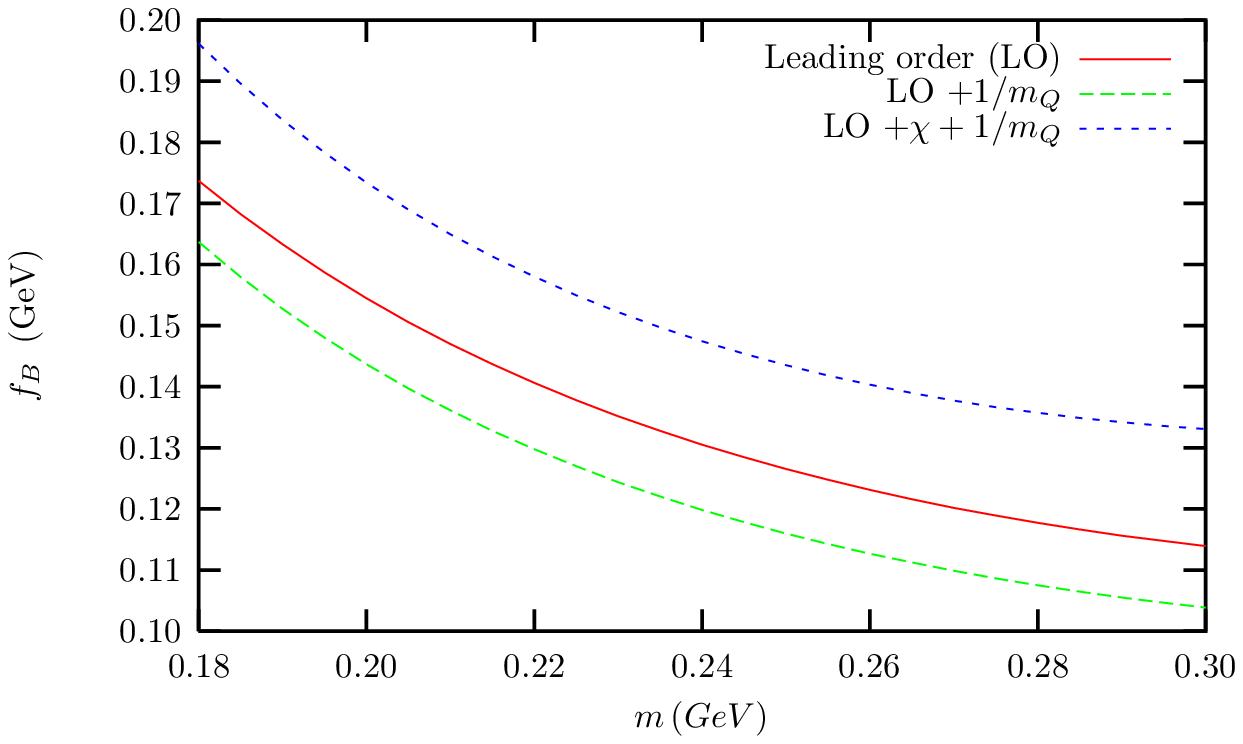}
\caption{$f_B$ as a function of $m$}
\label{fig:Fb}
\end{center}
\end{figure}

The weak decay constants of heavy mesons have been calculated by many
groups. Typical results from lattice calculations are $f_B=(200\pm 30)$ MeV
 and $f_D = (225\pm 30)$ MeV 
\cite{bernard}. QCD sum rules gives $f_B= (180\pm 30)$ MeV 
and $f_D=(190 \pm 20)$ MeV
\cite{alexander2} and NRQCD gives 
$f_B=(147(11)^{+8}_{-12}(9)(6)\,$ MeV\cite{kahn}.
In order to constrain our parameters we will use 
the following combinations from QCD sum rules which have been evaluated rather
accurately\cite{beyalev}~:
\bea
f_B\,f_{B^*}\sqrt{2}\,\ga^{B^*B\pi}\,M_B/f_\pi\,=\,(0.64\,\pm\,0.06)\, \text{GeV}^2
 \; ,\nonumber \\
f_D\,f_{D^*}\sqrt{2}\,\ga^{D^*D\pi}\,M_D/f_\pi\,=\,(0.51\,\pm\,0.05)\, \text{GeV}^2
\; , 	
\label{prod}
\eea  
where $\ga^{H^*H\pi}$ (for $H=B,D$) is the chiral coupling $\ga$ with chiral
 and $1/m_Q$ corrections included. The left-hand side of (\ref{prod}) - for 
$H=B,D$ respectively- is a function of $m$ and (the uncorrected) $\ga$.
Therefore (\ref{prod}) gives $\ga$ as a function of the constituent 
light quark mass $m$, which has been plotted in figure \ref{fig:Ga}
and \ref{fig:Gac}. 
\begin{table}[t]
\begin{center}
\begin{tabular}{l c}\hline\hline
\multicolumn{2}{c}{Input parameters} \\ \hline
$m$             & $(220\pm 30)$ MeV \\
$f_\pi$           & $93\,$ MeV \\
$\qc$           & $-(0.240\,$ GeV$)^3$ \\
$\alpha_s(\Lambda_\chi)$ & 0.50 \\
$\mu_G^2(B)$        & $0.36\,$ GeV$^2$ \\
$f_Bf_{B^*}\sqrt{2}M_B\tilde{g}^{B^*B\pi}_{\cal A}/f_\pi$ & 
\; \, $(0.64\pm0.06)\,$ GeV$^2$\\
$\alpha_s(m_b)$ & 0.21 \\
$m_b$             & 4.8 GeV \\
$C_\gamma^b(\Lambda_\chi)$ & 1.1 \\
$C_v^b(\Lambda_\chi)$ & 0.05 \\
$C_{M}^b(\Lambda_\chi)$ & 0.85\\
$C_{K}^b(\Lambda_\chi)$ & 0.25\\
$\mu_G^2(D)$        & $0.30\,$ GeV$^2$ \\
$f_Df_{D^*}\sqrt{2}M_D\tilde{g}^{D^*D\pi}_{\cal A}/f_\pi$ & 
\; \, $(0.51\pm0.05)\,$ GeV$^2$\\
$\alpha_s(m_c)$ & 0.36 \\
$m_c$             & 1.4 GeV \\
$C_\gamma^c(\Lambda_\chi)$ & 0.9 \\
$C_v^c(\Lambda_\chi)$ & 0.08 \\ 
$C_{M}^c(\Lambda_\chi)$ & 1.15\\
$C_{K}^c(\Lambda_\chi)$ & 0.75\\ \hline   \hline\hline
\end{tabular}
\end{center}
\caption{Input parameters of HL$\chi$QM in the $B$- and $D$- sector.}

\label{tab:predicted}
\end{table}
Thus, we can then plot $f_B$, $f_D$ and other quantities as a function
 of the  light quark mass. We may further use explicit values for
 $f_B$ and  $f_D$ to determine a value for $m$ in the $B$- and $D$- meson
sector separately. However, because we consider the Isgur-Wise function
which involve both $B$- and $D$- mesons, we need a unique value of $m$,
which give a reasonable value of $f_B$ and $f_D$ simultaneously.
As can be seen from the plot in figure \ref{fig:Fb}, 
this can be accomplished by taking
\begin{equation}
m\,=\, (220 \pm\ 30)\,\text{MeV} \; .
\end{equation}
This is consistent with the value $m= (200\pm 5)$ MeV used \cite{BEF} in the
pure light sector in order to fit the $\Delta I = 1/2$ rule for
$K \rightarrow 2 \pi$ decays. 


 \begin{figure}[t]
\begin{center}
   \epsfig{file=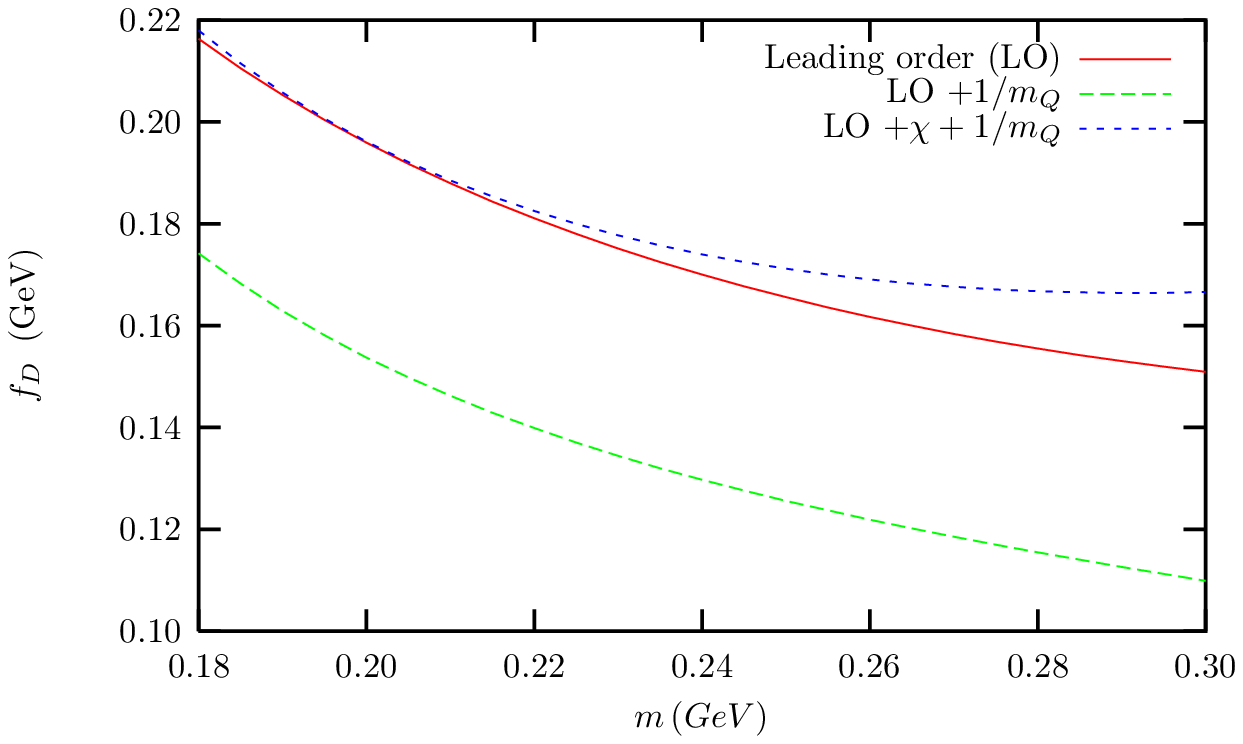}
\caption{$f_D$ as a function of $m$}
\label{fig:Fd}
\end{center}
\end{figure}
\begin{table}[t]
\begin{center}
\begin{tabular}{l l}\hline\hline
\multicolumn{2}{c}{Predictions of $HL\chi QM$} \\ \hline
$G_B$           & $\,(7.7\pm 0.6)\,$ GeV$^{-1/2}$ \\
$\gc^{1/4}$     & $\,(0.315\pm 0.020)\,$ GeV \\
$g_1$           & $(1.3\pm 0.2)\,$ GeV \\
$g_2$           & $(0.39\pm 0.05)\,$ GeV \\
$\e_1$          & $-(0.7\pm 0.2)\,$ GeV \\
$\lambda_1$     & $\,1.0\pm 0.2$\\
$\mu_\pi^2$     & $\,(0.41\pm 0.02)\,$ GeV$^2$ \\
$g_{\cal A}$    & $\,(0.42\pm 0.06)$ \\
$g^{B^*B\pi}_{\cal A}$    & $\,(0.31\pm 0.11)$ \\
$g^{B^*B^*\pi}_{\cal A}$    & $\,(0.22\pm 0.13)$ \\
$f_B$           & $\,(165\pm 20)\,$ MeV \\
$f_{B^*}$       & $\,(170\pm 25)\,$ MeV \\
$f_{B_s}$       & $\,(170\pm 20)\,$ MeV \\
$f_{B^*_s}$     & $\,(175\pm 25)\,$ MeV \\
$f_{B^*}/f_B$  & $\,1.06\pm 0.03$	\\
$f_{B_s}/f_B$   & $\,1.07\pm 0.02$ \\
\hline\hline
\end{tabular}
\end{center}
\caption{Predictions of HL$\chi$QM in the $B$- sector, the errors
in the predictions is a consequence of the error bars in the input
parameters.}\label{tab:predictedb}
\end{table} 
\begin{table}[t]
\begin{center}
\begin{tabular}{l l}\hline\hline
\multicolumn{2}{c}{Predictions of HL$\chi$ QM} \\ \hline
$G_D$           & $(6.8\pm 0.4)\,$ GeV$^{-1/2}$ \\
$\gc^{1/4}$     & $(0.300\pm 0.020)\,$ GeV \\
$g_1$           & $ (0.8\pm 0.1)\,$ GeV \\
$g_2$           & $  (0.33\pm 0.04)\,$ GeV \\
$\e_1$          & $-(0.6\pm 0.2)\,$ GeV \\
$\lambda_1$     & $0.7\pm 0.1$\\
$\mu_\pi^2$     & $(0.32\pm 0.03)\,$ GeV$^2$ \\
$g_{\cal A}$    & $0.55\pm 0.08$ \\
$g^{D^*D\pi}_{\cal A}$ & $0.46\pm 0.15$ \\
$g^{D^*D^*\pi}_{\cal A}$    & $(0.27\pm 0.22)$ \\
$f_D$           & $(190\pm 20)\,$ MeV \\
$f_{D^*}$       & $(220\pm 35)\,$ MeV \\
$f_{D_s}$       & $(205\pm 15)\,$ MeV \\
$f_{D^*_s}$     & $(235\pm 35)\,$ MeV \\
$f_{D^*}/f_D$   & $1.22\pm 0.09$	\\
$f_{D_s}/f_D$   & $1.16\pm 0.03$ \\
\hline\hline
\end{tabular}
\end{center}
\caption{Predictions of HL$\chi$QM in the $D$- sector, the errors
in the predictions is a consequence of the error bars in the input
parameters.}\label{tab:predictedd}
\end{table}
 \begin{figure}[t]
\begin{center}
   \epsfig{file=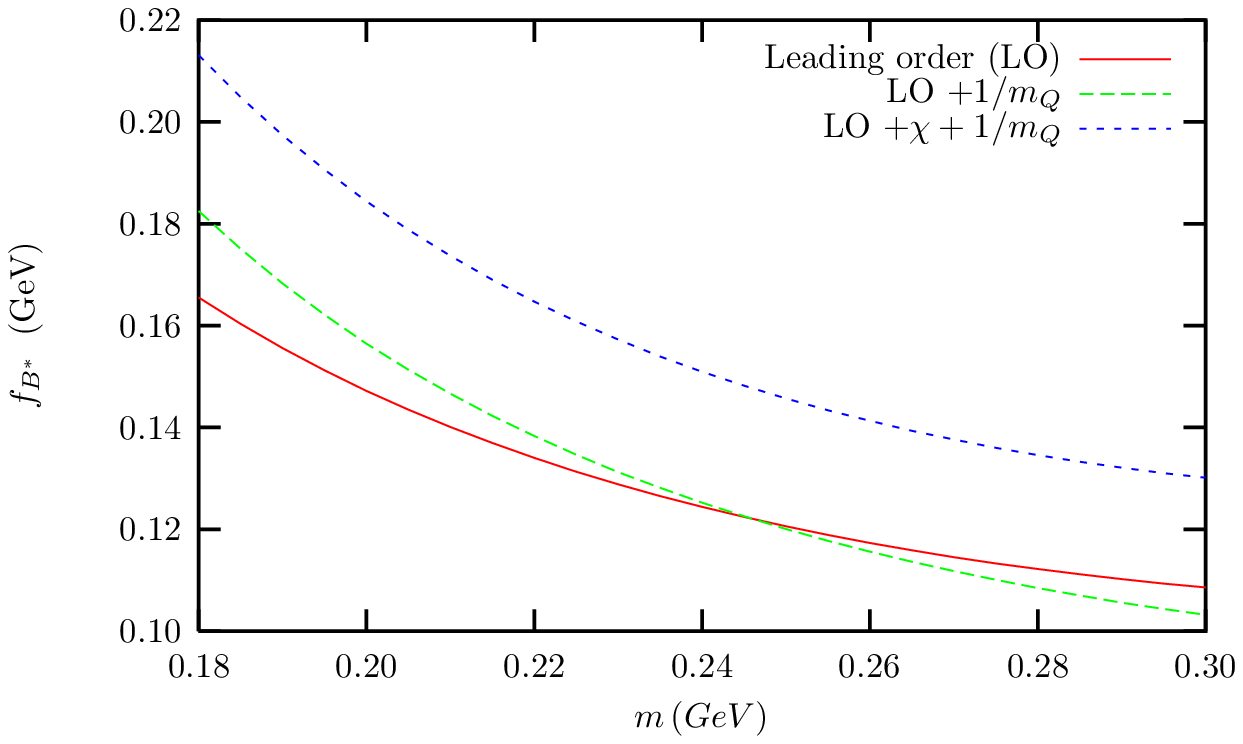}
\caption{$f_{B^*}$ as a function of $m$}
\label{fig:Fbs}
\end{center}
\end{figure}

\begin{figure}[t]
\begin{center}
   \epsfig{file=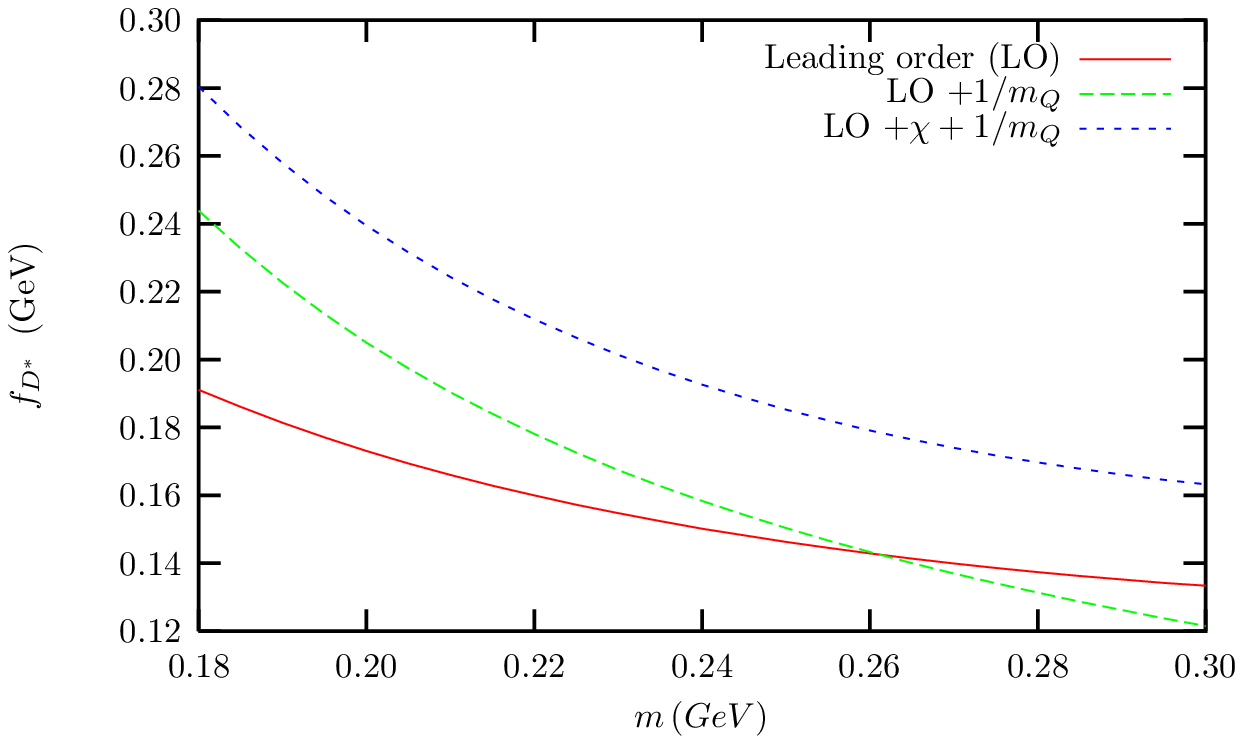}
\caption{$f_{D^*}$ as a function of $m$}
\label{fig:Fds}
\end{center}
\end{figure}

In table \ref{tab:predictedb} and table \ref{tab:predictedd}  we have
listed some of the predictions of HL$\chi$QM for decay constants and
counterterms. The input parameters have been listed in table
 \ref{tab:predicted}. It could be argued that the bare parameters,
$\ga,\, G_H,\,\gc$,
listed in table \ref{tab:predictedb} and table \ref{tab:predictedd}
should be equal. As statet earlier in this model $m$ has to have an unique walue
in both the $D$- and $B$-sector. In order to fit our model to the result from QCD
sumrules (equation (\ref{prod})), we have to alow for a different walue of
$\ga,\, G_H,\,\gc$ in the $B$- and $D$-sector. From table
\ref{tab:predictedb} and table \ref{tab:predictedd} we see
that these parameters agrees within errorbars in the two sectors.

The inclusion of the counterterm for $f_{B_s}$
is crucial, putting $\omega_1=0$ would give
$f_{B_s}/f_{B}=1.29\pm0.03$ and $f_{D_s}/f_{D}=1.34\pm0.04$  
which is much higher than most lattice estimates\cite{flynn}
 $f_{B_s}/f_{B}\simeq f_{D_s}/f_{D}\simeq 1.15$
 and QCD sumrules 
\cite{narison} $f_{B_s}/f_{B}=1.16\pm0.05$ and
$f_{D_s}/f_{D}=1.15\pm0.04$. From table \ref{tab:predictedb} and
\ref{tab:predictedd}, we see that we are a little low in the 
$B$-sector but in the $D$-sector the result agree nicely.
The decay constant $f_{D_s}$ have been measured \cite{rembold}
$f_{D_s}\,=\,(264\pm15\pm33\pm2\pm4)$ MeV and \cite{pdg}
$f_{D_s}\,=\,(280\pm19\pm28\pm34)$ MeV. It is somewhat higher than our
result, but we are within 1$\sigma$ of the experimental result.
The ratio $f_{H^*}/f_H$ has been
calculated in HQEFT sum rules with the result \cite{neub3}
 $f_{B^*}/f_B\simeq 1.07\pm0.03$ and 
$f_{D^*}/f_D\simeq 1.37\pm0.04$, which also agrees perfect with our
results.  
  
The coupling $\ga^{D^*D\pi}$ has been measured \cite{anastassov}
 $\ga^{D^*D\pi}\,=\,0.59\pm0.01\pm0.07$. Our prediction agrees well with this
 result.  The experimental result has also been predicted by a 
bag model calculation \cite{ahhh}, $\ga^{D^*D\pi}=0.60$ 
and $\ga^{B^*B\pi}=0.57$. 

In conclusion, we have constructed a model which gives a reasonably good 
description of   decay constants,
the chiral axial coupling $\ga$, masses and the Isgur-Wise function.
We observe that the coupling $\ga$ (leading order and corrected) is smaller 
in the $B$-sector than for the $D$-sector. Furthermore, we find that
$\mu_\pi^2 > \mu_G^2$ both in the $B$- and  the $D$-sector \cite{bigi}.  
We have also showed that it is possible to systematically calculate 
the $1/m_Q$ corrections. For the decay constants 
in both the $B$- and $D$-sector they are of
the same size as the chiral corrections as,
can be seen from figure \ref{fig:Fb}-\ref{fig:Fds}.

The model may be used to give predictions for other quantities. Especially, 
it will be suitable for calculation of the $B$-parameter for $B-\bar{B}$
mixing \cite{AHJOE}.   
\begin{acknowledgments}
This work was supported by The Norwegian Research Council 
\end{acknowledgments}

\appendix

\section{Loop integrals}\label{app:loop}

The divergent integrals entering in the bosonization of the 
HL$\chi$QM are defined :
\bea
I_1 \, &\eq &\,\int\fr{d^dk}{(2\pi)^d}\fr{1}{k^2 -  m^2} \\
I_{3/2}\, &\eq &\, \int\fr{d^dk}{(2\pi)^d}\fr{1}{(v\cdot k)(k^2 -  m^2)} \\
I_2\, &\eq\, &\int\fr{d^dk}{(2\pi)^d}\fr{1}{(k^2 -  m^2)^2} 
\eea

The $\kappa_i$'s  are defined as:
\begin{eqnarray}
&&\kappa_0 \eq - i\fr{m}{16\pi}\qquad
\widetilde{\kappa_0}\eq\,i\fr{m^2}{8\pi^2} \\
&&\kappa_1 \eq\, i\fr{8 -  3\pi}{384m^3}\qquad
\kappa_2 \eq\, i\fr{3\pi-4}{384m^2}\\
&&\kappa_3 \eq\, i\fr{8+ 3\pi}{384m^3}\qquad
\kappa_4 \eq\, i\fr{2-3\pi}{192m^2}\\
&&\kappa_5 \eq\, i\fr{1}{96m^2}
\; .
\eea

Integrals involving a heavy quark propagator and a light quark propagator
can be symmetried using the following formula :
\bea\label{georgi}
L^{m,\Delta}_{p,q}&&
\eq\int\,\fr{d^dk}{(2\pi)^d}
\fr{1}{(k^2 - m^2)^p(v\cdot k-\Delta)^q}\nonumber \\
&&=
\fr{\Gamma (p +  q)}{\Gamma (p)\Gamma(q)}
\int\limits_0^\infty\,d\lambda\int\,\fr{d^dk}{(2\pi)^d}\fr{2^q\lambda^{q -  1}}
{(k^2 -  m^2 +  2\lambda v\cdot k)^{p +  q}}
\eea
There are three interesting limits, where this integral can be written 
in a rather compact form : $\Delta, m\to 0$ and $\Delta=m$ :
\bea
L^{m,0}_{p,q}&&=2^{q-1}\fr{(-1)^{p+q}i}{(4\pi)^{d/2}}
\fr{\Gamma(q/2)\Gamma(p+q/2-d/2)}{\Gamma(p)\Gamma(q)}
\left(\fr{1}{m^2}\right)^{p+q/2-d/2}
\label{eq:L}\\
L^{0,\Delta}_{p,q}&&=\fr{(-1)^{p+q}i}{2^{2p-d}(4\pi)^{d/2}}
\fr{\Gamma(d/2-p)\Gamma(2p+q-d)}{\Gamma(p)\Gamma(q)}
\left(\fr{1}{\Delta^2}\right)^{p+q/2-d/2}\\
L^{m,m}_{p,q}&&=\fr{(-1)^{p+q}i}{(4\pi)^{d/2}}
\fr{\Gamma(p+q/2+1/2-d/2)
\Gamma(p+q/2-d/2)}{\Gamma(p+q+1/2-d/2)}
\left(\fr{1}{m^2}\right)^{p+q/2-d/2}\label{eq:Lmm}
\eea
where $d =  4 -  2 \epsilon$ is the dimension of space. As a check
equation (\ref{eq:L}) gets the well known form in the limit $q\to 0$ 
($\Gamma(q/2)/\Gamma(q)\rightarrow 2$),$L_{p,0}^{m,0}\eq I_p$. In the limit $p\to 0$,
$L^{m,0}_{0,q}=0$, 
this is because there is no mass scale entering in the integral. 

In the general case for $\Delta,m\neq 0$, there is no compact form of
$L_{p,q}^{m,\Delta}$, but all integrals needed in calculations 
can be relatet to the following integrals :
\begin{equation}\label{L11}
L_{1,1}^{m,\Delta}=\fr{-i}{8\pi}\left(
\fr{1}{\ebar}-\ln(m^2)+2-2F(m/\Delta)\right)
\end{equation}
\bea
&&\int\,\fr{d^dk}{(2\pi)^d}
\fr{k^{\mu}k^{\nu}}{(k^2 - m^2)(v\cdot k-\Delta)}=
A\,g^{\mu\nu}+B\,v^{\mu}v^{\nu}\nonumber \\
&&A=\fr{1}{d-1}\int\,\fr{d^dk}{(2\pi)^d}
\fr{k^2-(v\cdot k)^2}{(k^2 - m^2)(v\cdot k-\Delta)}\nonumber\\
&&=\fr{i\Delta}{16\pi^2}\left\{(-\fr{1}{\ebar}+
\ln(m^2)-1)(m^2-\fr{2}{3}\Delta)-\fr{4}{3}F(m/\Delta)(\Delta^2-m^2)
-\fr{4}{3}(m^2-\fr{5}{6}\Delta^2)\right\}\\
&&B=-A+\int\,\fr{d^dk}{(2\pi)^d}
\fr{(v\cdot k)^2}{(k^2 - m^2)(v\cdot k-\Delta)}\nonumber\\
&&=\fr{-i\Delta}{16\pi^2}\left\{(-\fr{1}{\ebar}+
\ln(m^2)-1)(2m^2-\fr{8}{3}\Delta)-\fr{4}{3}F(m/\Delta)(4\Delta^2-m^2)
\right.\nonumber\\ &&\left.\qquad\qquad
-\fr{4}{3}(m^2-\fr{7}{3}\Delta^2)\right\}\label{Lmunu}
\eea
where:
\begin{equation}\label{F}
F(x)=\begin{cases}
&-\sqrt{x^2-1}\tan^{-1}(\sqrt{x^2-1})\qquad x>1\\
&\,\sqrt{1-x^2}\tanh^{-1}(\sqrt{1-x^2})\qquad x<1
\end{cases}
\end{equation}
In the case of $\Delta>m$ we have ignored an analytic 
real part in (\ref{L11}). As a check on these calculations we can use
equation (\ref{eq:Lmm}). Equation (\ref{L11}) coincides with the one
obtained in \cite{GriBo} however equation (\ref{Lmunu}) differs
by a factor $-2/3(m^2-2/3\Delta^2)$ inside the parenthesis of the
expressions for $A$ and $B$. This is presumably due to the factor
$1/(d-1)=(1-2/3\e)/3$ in $A$.  

In the case of the IW function we also need the following integral :
\bea
L(p,\omega)&&\eq \int\,\fr{d^dk}{(2\pi)^d}\fr{1}{(v\cdot
k)(v^\prime\cdot k)(k^2-m^2)^p}\nonumber \\
&&=
\fr{2i(-1)^p}{(4\pi)^{d/2}}\fr{\Gamma(p+1-d/2)}{\Gamma(p)(m^2)^{p+1-d/2}}r(\omega)\quad
\text{where}\nonumber \\
r(\omega)&&\eq\fr{1}{\sqrt{\omega^2-1}}\, \text{ln}\left(\omega+\sqrt{\omega^2-1}\right)
\eea

\section{$SU(3)$ transformation Properties}\label{app:tranf}

The Lagrangian of the light quark sector is  
\begin{equation}
{\cal L}_{\chi QM} =   \bar{q}_L i \gamma \cdot D \, q_L \,  +   \, 
\bar{q}_R i \gamma \cdot D \, q_R
 -   \bar{q}_L {\cal M}_q \, q_R \,  -    \bar{q}_R {\cal M}_q^\dagger \, q_L
  -     m(\bar{q}_R \Sigma^{\dagger} q_L   +    \bar{q}_L \Sigma q_R) \; , 
\label{CQMLR}
\end{equation}
 The left -   and right -  handed projections of the quark fields transform as
\begin{equation}
q_L \rightarrow V_L \, q_L \quad , \qquad q_R \rightarrow V_R \, q_R \; ,
\end{equation}
where $V_L \, \epsilon \, SU(3)_L$ and $V_R \, \epsilon \, SU(3)_R$.
 The octet meson field $\Sigma$ transforms as
\begin{equation}
 \Sigma  \rightarrow V_L \, \Sigma  \, V_R^\dagger \; .
\end{equation}
The $\xi$ field transforms more complicated as
\begin{equation}
 \xi  \rightarrow U  \, \xi  \, V_R^\dagger \,  =    V_L \xi U^\dagger
\end{equation}
where $U \epsilon SU(3)_V$. The constituent quark fields 
$\chi_L  =   \xi^{\dagger} q_L$ and $\chi_R  =   \xi q_R$, and the heavy meson field 
transform in a simple way under $SU(3)_V$:
\begin{equation}
\chi_L \rightarrow U \, \chi_L \quad , \qquad \chi_R \rightarrow U \; \chi_R
\quad , \qquad   H_{v} \rightarrow  H_{v} \, U^\dagger \; .
\end{equation}
The vector and axial fields transform as 
\begin{equation}
{\cal V}_{\mu} \rightarrow U \, {\cal V}_{\mu} \, U^\dagger \;  +   \; 
i U \partial_\mu \, U^\dagger \quad , \qquad 
{\cal A}_\mu \rightarrow  U {\cal A}_\mu \, U^\dagger \; . 
\end{equation}
Note that the weak current in (\ref{Lcur}) transforms as
\begin{equation}
J_f^\alpha \rightarrow J_h^\alpha \left( V_L^\dagger \right)_{h f} \; .
\end{equation}


\bibliographystyle{unsrt}

\end{document}